\documentclass[prl,amsmath,amssymb,reprint,superscriptaddress]{revtex4-1}
\usepackage[usenames]{color}

\usepackage{times}
\usepackage{graphicx}
\usepackage{mhchem}

\usepackage[labelfont=bf]{caption}
\usepackage{tikz}

\begin{document}

\date{\today}

\title{Dynamic Symmetry Breaking and Spin Splitting in Metal Halide Perovskites}

\author{Scott McKechnie}
\affiliation{Department of Physics, King's College London, London WC2R 2LS, UK}

\author{Jarvist M. Frost}
\affiliation{Department of Materials, Imperial College London, London SW7 2BZ,UK}

\author{Dimitar Pashov}
\author{Pooya Azarhoosh}
\affiliation{Department of Physics, King's College London, London WC2R 2LS, UK}

\author{Aron Walsh*}
\affiliation{Department of Materials, Imperial College London, London SW7 2BZ,UK}
\affiliation{Department of Materials Science and Engineering, Yonsei University, Seoul 03722, Korea}
\email{a.walsh@imperial.ac.uk}

\author{Mark van Schilfgaarde*}
\affiliation{Department of Physics, King's College London, London WC2R 2LS, UK}
\email{mark.van\_schilfgaarde@kcl.ac.uk}

\begin{abstract}
Metal halide perovskites exhibit a materials physics that is distinct from traditional inorganic and organic semiconductors. 
While materials such as \ce{CH3NH3PbI3} are non-magnetic, the presence of heavy elements (Pb and I) in a 
non-centrosymmetric crystal environment result in a significant spin-splitting of the frontier electronic bands through the
Rashba-Dresselhaus effect. We show, from a combination of \textit{ab initio} molecular dynamics, density-functional 
theory, and relativistic quasi-particle \textit{GW} theory, that the nature (magnitude and orientation) of the band 
splitting depends on the local asymmetry around the Pb and 
I sites in the perovskite structure. The potential fluctuations vary in time as a result of thermal disorder 
and a dynamic lone pair instability of the Pb(II) 6s$^{2}$6p$^{0}$ 
ion. We show that the same physics emerges both for the organic-inorganic \ce{CH3NH3PbI3} and the 
inorganic \ce{CsPbI3} compound. The results are relevant to the 
photophysics of these compounds and are expected to be general to other lead iodide containing perovskites.
\end{abstract}

\maketitle

In crystals that preserve both time-reversal and spatial-inversion symmetry, 
there is at least a double degeneracy of the electron energy bands $E_{n}(k)$ for 
each wave vector $k$ in the Brillouin zone \cite{Elliott1954a}. However, the removal of             
spatial inversion symmetry produces an asymmetric potential,                                
which, through the spin-orbit interaction, lifts the two-fold spin degeneracy \cite{Winkler2003}. 
Dresselhaus and Rashba were the first to provide group theory descriptions of the 
effect of spin-orbit coupling (SOC) on non-centrosymmetric zincblende \cite{Dresselhaus1955} and 
wurtzite \cite{Rashba1959,Rashba1959b,Rashba1960a} structures \cite{bihlmayer2015}. 
In a later work, Bychkov and Rashba pointed out that the analysis also applied to 2D 
systems \cite{Bychkov1984}. The so-called Rashba effect is associated with a spin splitting of the 
dispersion $E(k)$ that is linear in $k$ and a spin texture that                          
is helical \cite{Rashba1959b,Winkler2003,Winkler2004}. On the other hand, the Dresselhaus effect typically displays a splitting  
that is cubic in $k$ and a nonhelical spin texture \cite{Dresselhaus1955,Winkler2003,Winkler2004}. 
A recent study by Zunger et al. \cite{Zhang2014}, provides a concise definition of the Rashba and 
Dresselhaus spin splittings, emphasising the local origin of these effects: atomic sites 
without inversion symmetry can either be polar (local Rashba) or nonpolar (local Dresselhaus) 
and the bulk effect is a vector sum over all sites (allowing for the coexistence of 
bulk Rashba and bulk Dresselhaus). Both are rooted in spin-orbit interactions 
at sites with local asymmetry and result in spin-split bands; from here on, 
`spin splitting' will be used as a collective term for the two effects. 

Spin splitting is now seen as a crucial feature in the 
electronic structure of solid-state systems. Indeed, exploiting these effects to control 
the electron spin is a central goal in spintronics \cite{Zutic2004,Fabian2007}. It is also emerging 
as an important aspect of materials for photovoltaics, where it is thought to strongly influence the                 
absorption and transport properties in perovskite solar cells
\cite{Brivio2014,Kim2014,Amat2014,Motta2015,Kepenekian2015,Zheng2015,Etienne2016,Mosconi2017,Tan2017,Hutter2017,Niesner2016,Wang2017,Niesner2017}.
Metal halide perovskites have attracted significant attention 
for solar-energy conversion, with champion light-to-electricity
conversion efficiencies exceeding 22 \% for laboratory scale
devices \cite{Yang2017,Kojima2009}. 
The unique chemistry and physics of these materials is also extending their range of applications to 
areas including light-emitting diodes, spintronics, solid-state memory and
solid-state sensors \cite{Brenner2016,Ahmadi2017}.

The archetype material, \ce{CH3NH3PbI3 (or \ce{MAPbI3})}, is composed of corner-sharing 
lead-iodide octahedra, with a charge-balancing organic cation in the void regions, 
forming a three-dimensional perovskite structure. With the large nuclear charges 
of lead and iodine, the electron kinetic energies near these cores are large. 
Relativistic effects become important. 
These materials also exhibit reduced symmetry (both static and dynamic) resulting in 
local asymmetry which, through the Rashba and Dresselhaus effects, can 
split the bands into separate spin channels and shift the location of valence 
and conduction band extrema. This can lead to spin and momentum forbidden 
transtions that suppress electron-hole recombination \cite{Zheng2015,Motta2015,Azarhoosh2016}. 
First-principles electronic-structure calculations 
predicted that such relativistic effects could result in a direct to 
indirect bandgap transition in Pb-based perovskites
\cite{Brivio2014,Kim2014,Amat2014,Motta2015,Kepenekian2015,Zheng2015}.
There are increasing data to support this hypothesis. 
Indirect transitions are implicated for the observed slow minority carrier
recombination \cite{Hutter2017,Niesner2016,Wang2017,Niesner2017}. 
In particular, angle-resolved X-ray photoemission spectroscopy (ARPES) has been used to directly resolve
indirect features in the band structure \cite{Niesner2016}. 

In this Letter, we study the effect of local symmetry breaking (arising both 
statically and dynamically) on the electronic structure of 
lead halide perovskites. We sample structures from finite-temperature molecular 
dynamic simulations and evaluate the spin splitting. The band splitting directly 
impacts the photophysics of an operating solar cell, and it is also a relevant 
metric of local asymmetry.
By making identical studies of both organic-cation and inorganic perovskites,
we show that similar behaviour is present. 
Both organic and inorganic finite-temperature structures exhibit significant band splitting. 
The soft polar nature of halide perovskites leads to large distortions and 
resulting asymmetry at heavy atom sites. 
With Pb-based perovskites, this leads to particularly pronounced spin splittings. 
We correlate a metric relevant for
device physics (spin-split bands) directly with features that can be observed
(lattice distortion).

\begin{figure}
\centering
\begin{tikzpicture}
\node (a) {\includegraphics[trim={4cm 4cm 7cm 8cm},clip,width=0.3\linewidth]{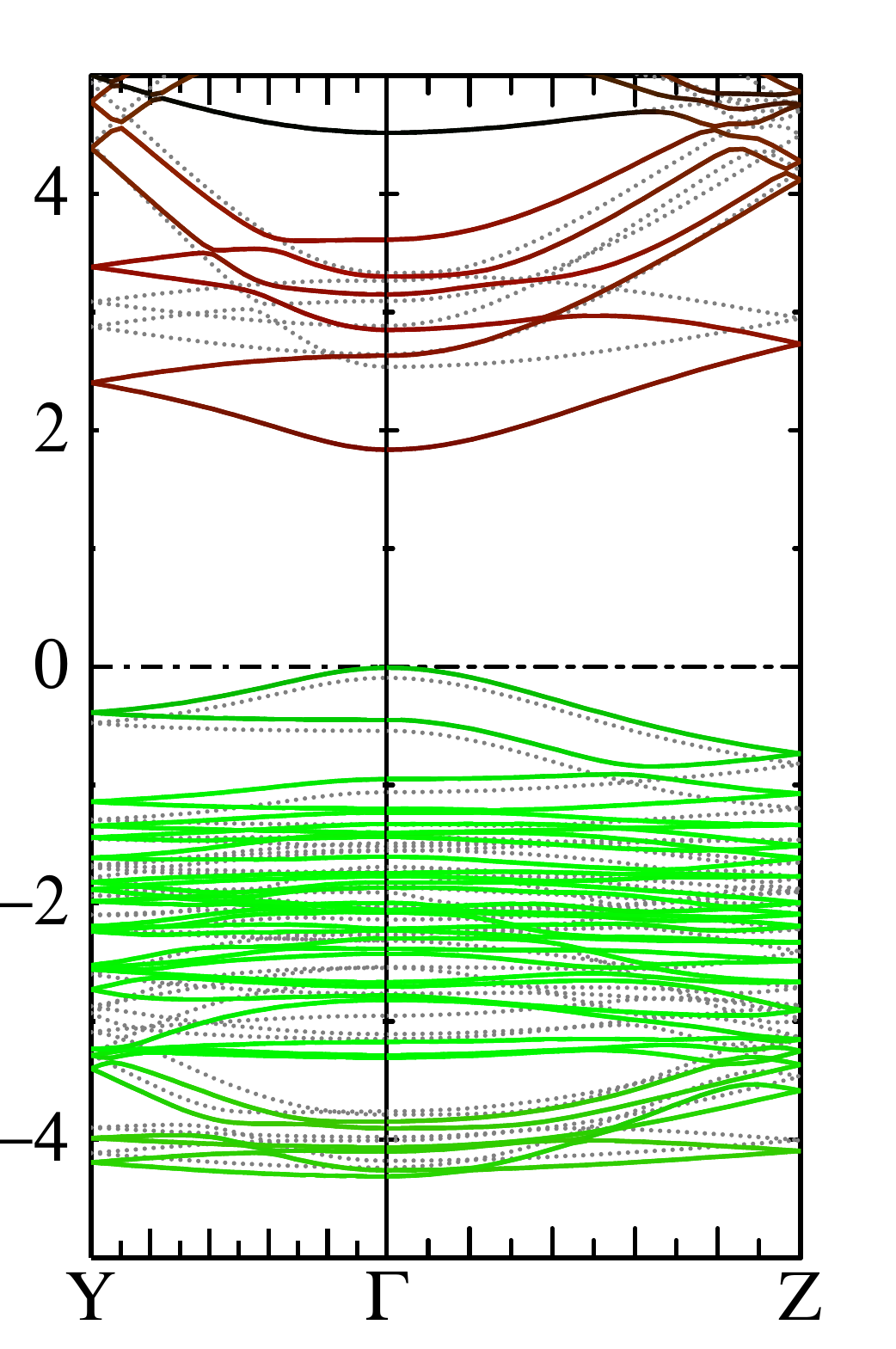}};
\node (b) [right of=a, yshift=-0.008cm, xshift=1.5cm] {\includegraphics[trim={4cm 4cm 7cm 8cm},clip,width=0.3\linewidth]{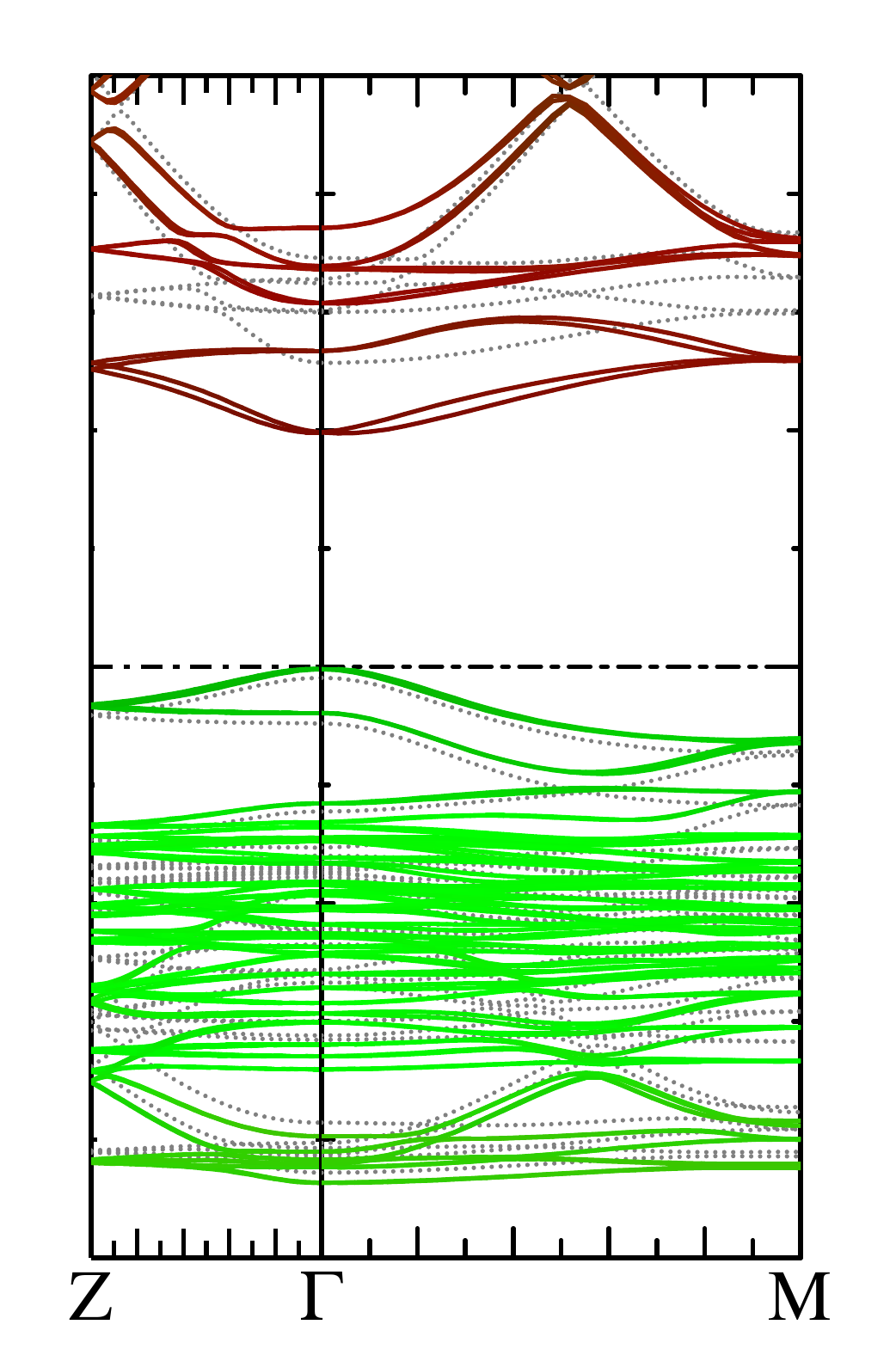}};
\node (c) [right of=b, yshift=-0.008cm, xshift=1.5cm] {\includegraphics[trim={4cm 4cm 7cm 8cm},clip,width=0.3\linewidth]{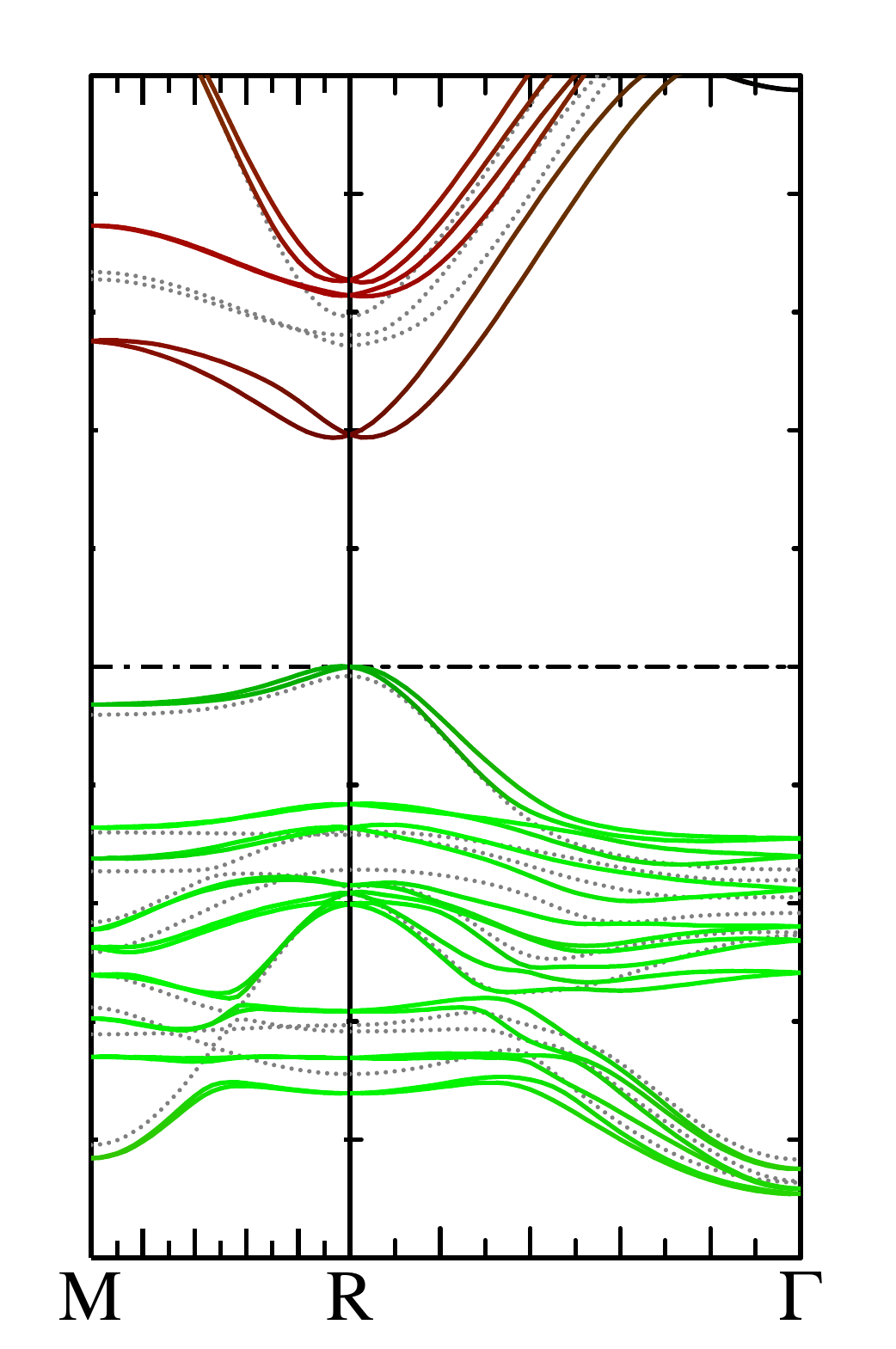}};
\node (d) [below of=a, yshift=-3cm, xshift=0cm] {\includegraphics[trim={4cm 4cm 7cm 8cm},clip,width=0.3\linewidth]{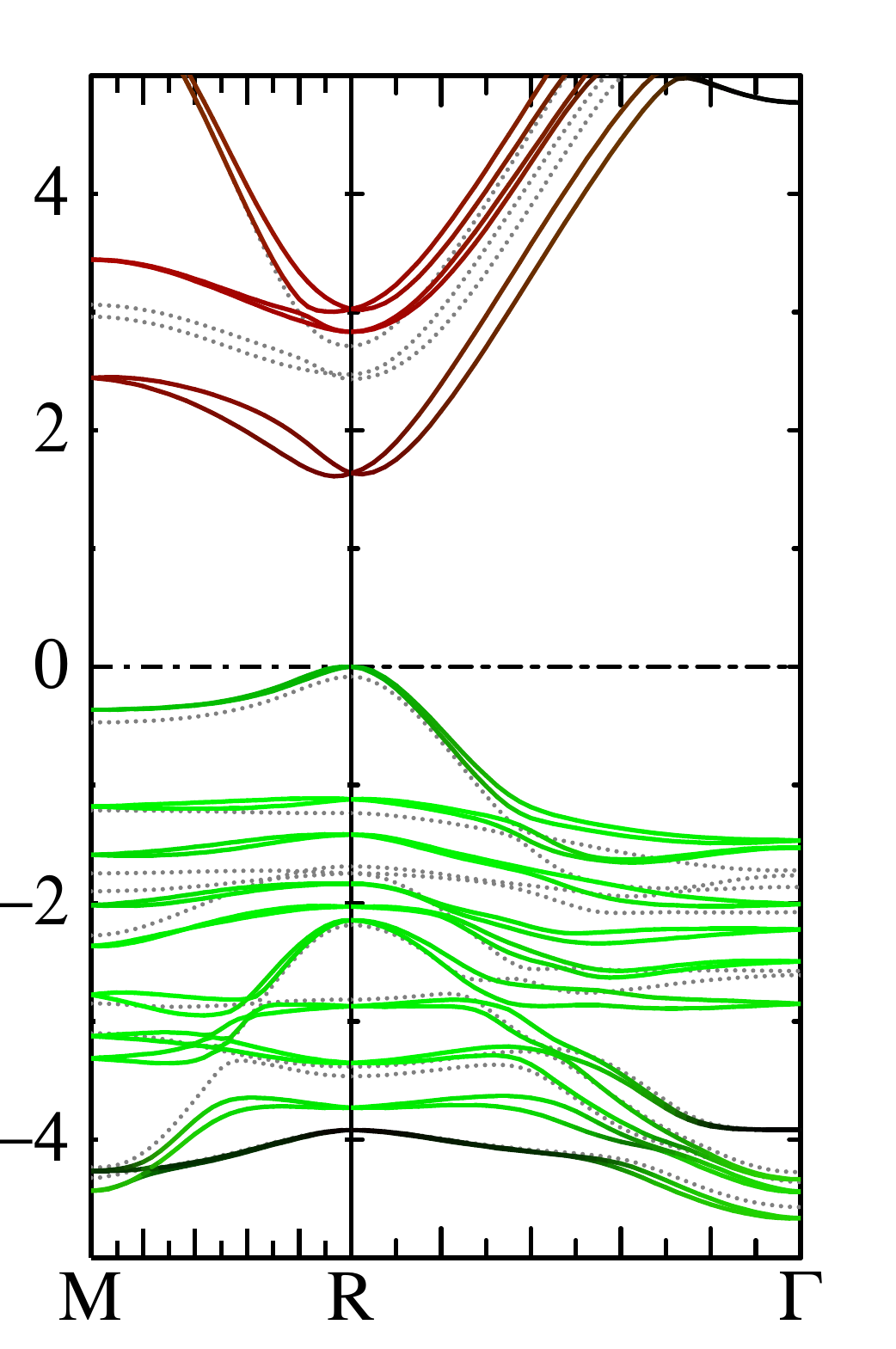}};
\node (e) [right of=d, yshift=-0.008cm, xshift=1.5cm] {\includegraphics[trim={4cm 4cm 7cm 8cm},clip,width=0.3\linewidth]{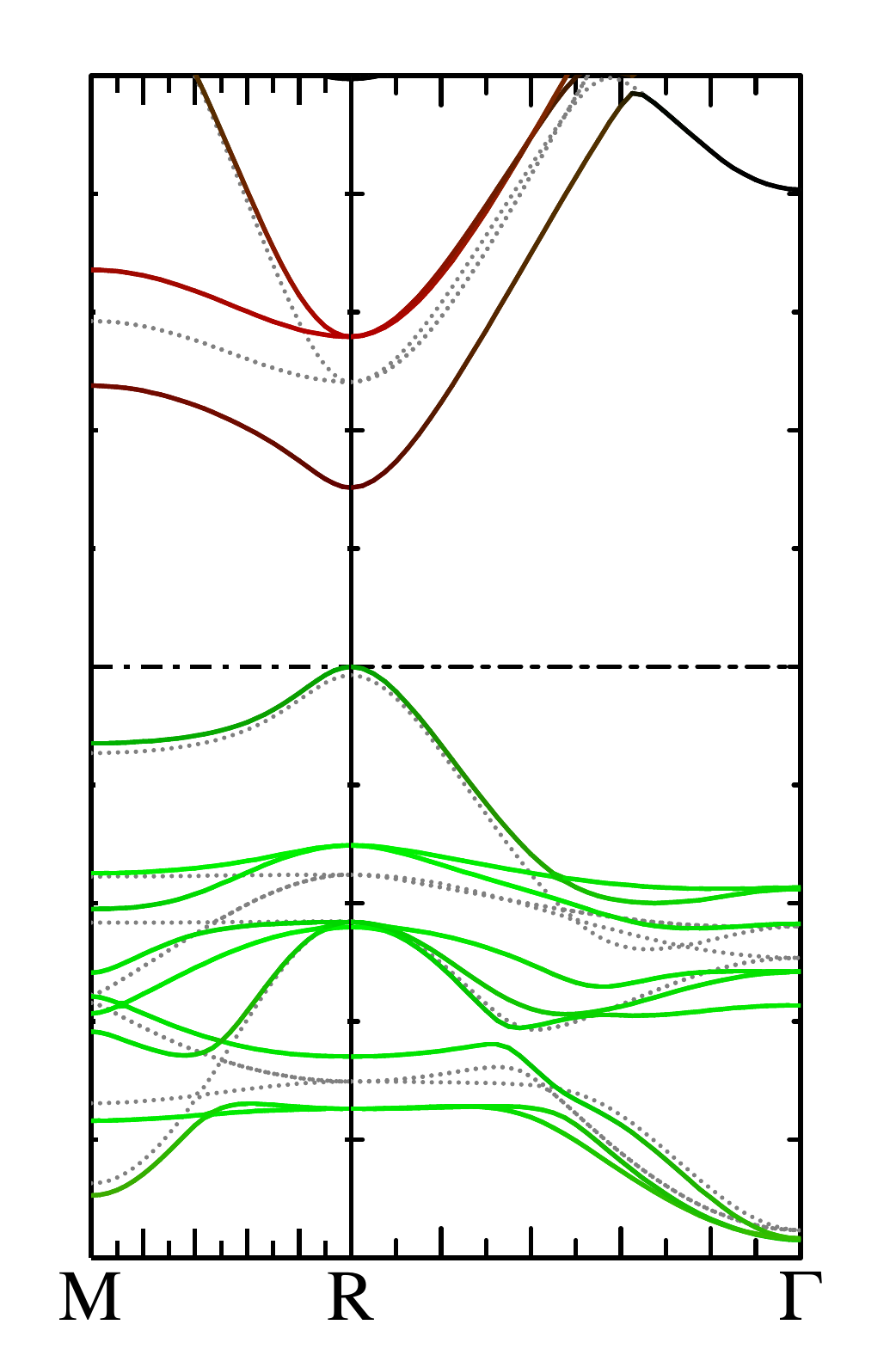}};
\node (f) [right of=e, yshift=-0.008cm, xshift=1.5cm] {\includegraphics[trim={4cm 4cm 7cm 8cm},clip,width=0.3\linewidth]{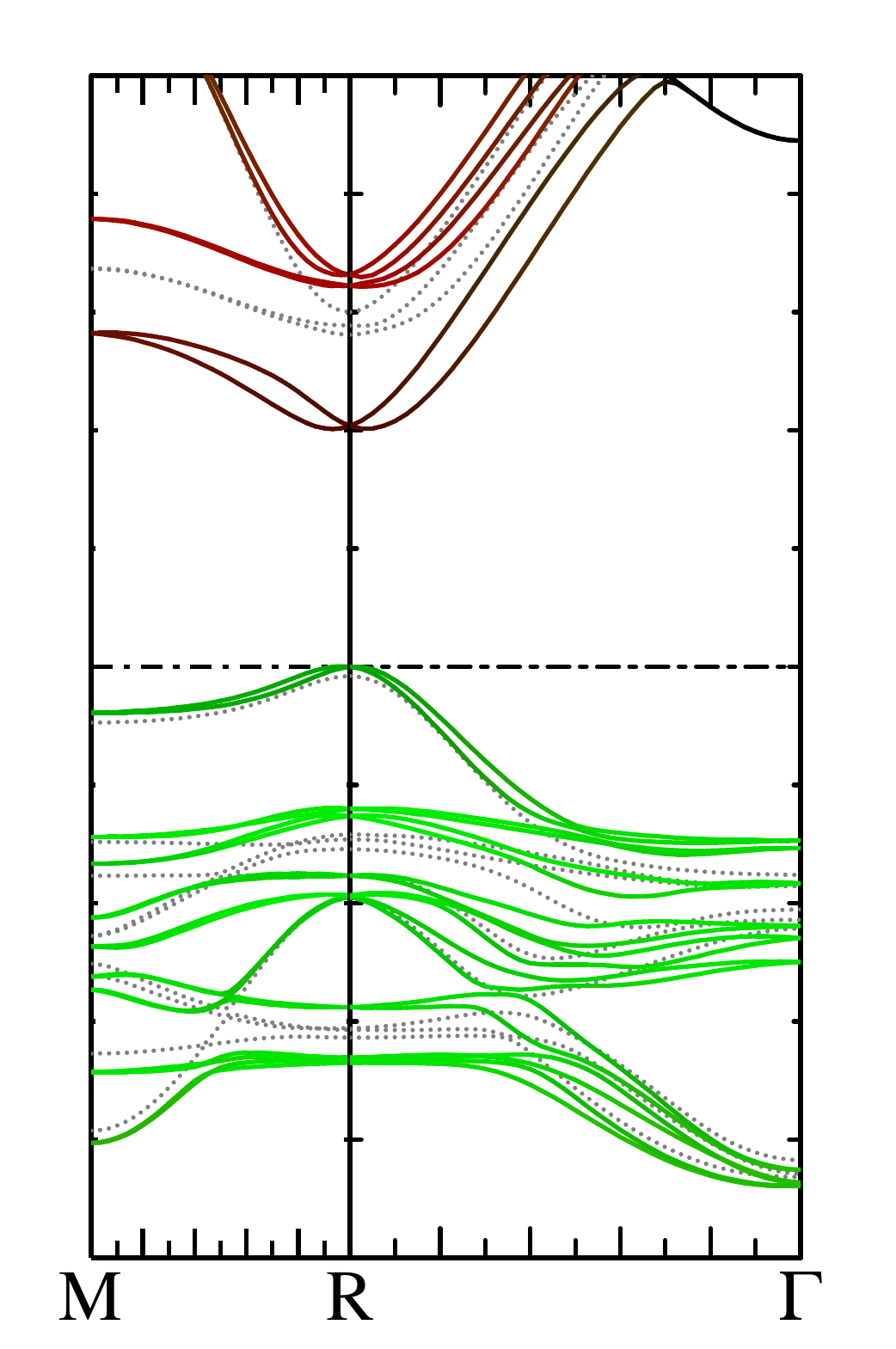}};
\node [above of=a, yshift=0.88cm] {\textbf{a}};
\node [above of=b, yshift=0.88cm] {\textbf{b}};
\node [above of=c, yshift=0.88cm] {\textbf{c}};
\node [above of=d, yshift=0.88cm] {\textbf{d}};
\node [above of=e, yshift=0.88cm] {\textbf{e}};
\node [above of=f, yshift=0.88cm] {\textbf{f}};
\node [left of=a, xshift=-0.5cm, rotate=90] {Energy (eV)};
\node [left of=d, xshift=-0.5cm, rotate=90] {Energy (eV)};
\end{tikzpicture}
\caption{Relativistic QS\textit{GW} electronic band structures.  \ce{CH3NH3PbI3} in its (a) orthorhombic, (b) tetragonal and (c) pseudocubic phases.
\ce{HC(NH2)2PbI3} in a (d) pseudocubic structure and CsPbI$_{3}$ in (e) cubic and (f) pseudocubic
structures. The pseudocubic CsPbI$_{3}$  is obtained by substituting Cs into \ce{CH3NH3PbI3}. Bands are coloured based on a Mulliken projection: Pb
p states are highlighted in red and I p states in green.}
  \label{f1}
\end{figure}
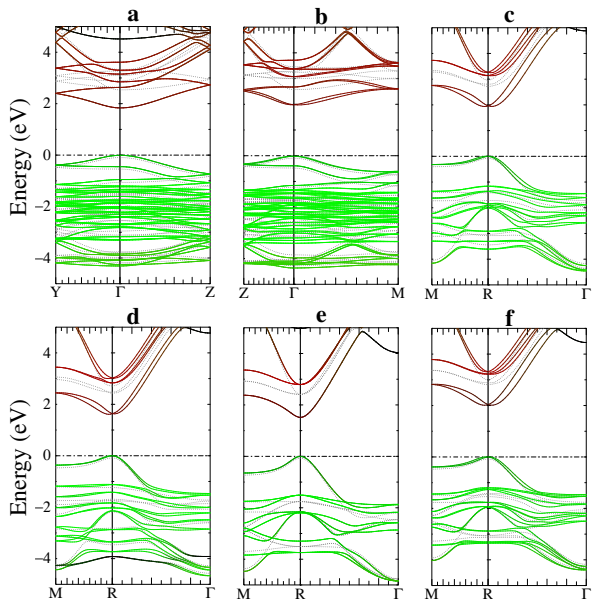


The electronic structure of lead halide perovskites is sensitive to the choice of 
Hamiltonian \cite{Whalley2017,Even2013}.
Relativistic effects lower and split the atomic levels of Pb and I (Table S1). 
These effects carry over into the 
condensed phase, resulting in a large renormalization of the bandgap (Table I). There is in 
addition, a $k$-dependent contribution originating from local asymmetry and spin-orbit 
coupling, which gives rise to the aforementioned spin splitting. As a result, the 
inclusion of scalar-relativistic and SOC corrections is essential for a correct description of the 
band structure \cite{Even2013,Brivio2014}. 
In these materials, the quasi-particle corrections are large, which necessitates
going beyond Kohn-Sham density-functional theory (DFT) \cite{Umari2014,Brivio2014,Filip2014}.
Here we use relativistic quasi-particle \textit{GW} theory (QS\textit{GW}), as
implemented in the all-electron \textsc{QUESTAAL} \cite{questaal} software
package. For its greater computational efficiency, we also use the local density 
approximation (LDA) \cite{kohn1965self} to calculate the electronic structure of structures 
sampled from molecular dynamics (MD) trajectories, and a set of 
structures sampled from lattice dynamics. 
These we reference against QS\textit{GW} calculations on individual cases. 
Further computational details are provided as Supplemental Material.

\textbf{Static crystal structure representations.}
As an initial reference we study static geometry-relaxed crystal structures. 
Our calculated QS\textit{GW} band structures of \ce{CH3NH3PbI3} in
three phases are shown in Fig. \ref{f1}. 
These structures have been carefully energy-minimized for lattice dynamics
calculations \cite{Brivio2015}. 
The low-temperature orthorhombic ($D_{2h}$ point group) phase has
a conventional-looking band structure with a direct bandgap. 
The principal effect of SOC (compare dashed grey and colored lines)
is a large renormalization of the conduction band energy, a result 
of the mainly Pb-derived $p$ level splitting into $p_{1/2}$ and $p_{3/2}$ states 
(see Table 1 and Fig. S1). 
The bandgap remains direct but decreases in magnitude by circa 0.8 eV. 

The centre of inversion present in the orthorhombic phase of \ce{CH3NH3PbI3}
is lost in static representations of the tetragonal and cubic phases. 
There is now a shift of the band extrema away from the high-symmetry points: 
spin splitting is activated. 
The indirect bandgaps are 2.00 eV and 1.87 eV for the tetragonal and pseudocubic phases,
and conduction-band minima are offset by 0.03 \AA$^{-1}$ and 0.06 \AA$^{-1}$
from high symmetry points. 
The latter calculated value is a giant spin splitting, comparable in magnitude to the 
largest values experimentally observed in bulk systems \cite{Ishizaka2011}. These band 
splitting features are consistent with the electron-hole 
recombination kinetics observed in experiment. In particular, there is a sharp fall off in the 
charge carrier lifetime at 
the transition to the orthorhombic phase, where a direct gap is recovered \cite{Hutter2017,Wang2017}.

The inversion-asymmetric tetragonal and pseudocubic structures have a Rashba-like 
spin splitting, with the usual $k$-linear dispersion relation and helical spin texture \cite{Kim2014,Kepenekian2015,Zheng2015}. 
However, instead of a ring of extrema there are two distinct extremal points. These are 
antipodal to each other, with a connecting path that passes through two saddle points \cite{Brivio2014}. 
At the DFT level, the conduction and valence bands have approximately 
the same momentum offset $\Delta k$ (difference in momentum between high-symmetry point and extremal point). 
However, QS\textit{GW} results show that $\Delta k$ in the valence band (VB) is 
roughly half that of the conduction band (CB). Relative to QS\textit{GW}, LDA-DFT also tends to overestimate 
the energy offset $\Delta E$ (defined as the difference in energy between the spin-split extremal point 
and high-symmetry point) in both bands. Similar trends are seen for other cations, 
such as FAPbI$_{3}$ and CsPbI$_{3}$. 
 
\begin{table}
\caption{\label{t1} Bandgaps (E$_g$) and conduction band splittings ($\Delta E$ and $\Delta k$) for a range of 
halide perovskites calculated
for static (athermal) structures using QS\textit{GW} including scalar-relativistic effects, and spin-orbit coupling
(SOC) where stated.
Measured bandgaps, E$_g$\textsuperscript{EXP}, are also shown.
}
\begin{ruledtabular}
\begin{tabular}{cccccc}
Material                  &   E$_g$ (eV)  &   E$_g$\textsuperscript{SOC}  & E$_g$\textsuperscript{EXP} &  $\Delta E$ (meV)   & $\Delta k$  (\AA$^{-1}$) \\
\hline
\ce{o\textendash MAPbI3}     &  2.63             &  1.84      &   1.65  \cite{Galkowski2016} &            0             &          0                  \\
\ce{t\textendash MAPbI3}     &  2.67             &  1.99      &    1.61  \cite{Galkowski2016} &            10            &          0.03               \\
\ce{pc\textendash MAPbI3}    &  2.72             &  1.87      &   1.63 \cite{Quarti2016} &            82            &          0.06               \\
\ce{pc\textendash FAPbI3}    &  2.52             &  1.61      &    1.52  \cite{Galkowski2016}  &            24            &          0.03               \\
\ce{pc\textendash CsPbI3}      &  2.81             &  1.95      &      &            82            &          0.06               \\
\ce{c\textendash CsPbI3}       &  2.48             &  1.51      &    1.73 \cite{Eperon2015}  &            0             &          0                  \\
\end{tabular}
\end{ruledtabular}
\end{table}
%
%
%

A central question is to what extent the calculated
spin splitting depends on the choice of cation. 
The methylammonium cation has a large built-in dipole \cite{Frost2014}, which
by itself applies a local electric field to the lead-iodide octahedra. 
Another common organic cation, formamidinium (\ce{HC(NH2)2} or FA), possesses a much
smaller dipole \cite{Frost2014} but still breaks the inversion symmetry of the unit cell, and applies
steric distortion to the octahedra (Fig. 1d). 
On the other hand, the inorganic cation in CsPbI$_{3}$ has spherical symmetry 
and there is no built-in dipole. The room-temperature phase is often assumed to be cubic ($O_h$
point group), with a direct gap (see Fig. 1e).

We consider these cations in a range of structures 
(Table \ref{t1} with band structures in Fig. 1).
Universally, once inversion symmetry is broken, a similar band splitting is observed.
The cations do not directly affect the band splitting. 
Instead, they are found to play an indirect role by influencing the lead-iodide octahedral distortions 
through steric effects; it is the cage deformations that generate the spin splitting. 
In the case of the inorganic cesium, the ideal cubic structure has no spin splitting. 
If we substitute cesium for methylammonium in the pseudocubic structure, and energy minimise 
with respect to cesium while keeping Pb and I fixed, then an almost identical spin 
splitting is generated (see Fig. 1f). 

In contrast to the average (long-range) cubic structure inferred from Bragg diffraction, 
recent measurements suggest that the structure of \ce{CsPbI3} is locally
symmetry broken at room temperature \cite{Bertolotti2017,Yaffe2017,Isarov2017}. 
Polar fluctuations have recently been observed in a number of other 
perovskites \cite{Guo2017}.
Anticipating that thermal disorder in these soft materials is necessary to understand the 
size and nature of the spin splitting, we now consider finite-temperature effects. 

\begin{figure}
\centering
\begin{tikzpicture}
\node (a) {\includegraphics[trim={0cm 0.5cm 0cm 0},clip,width=1\linewidth]{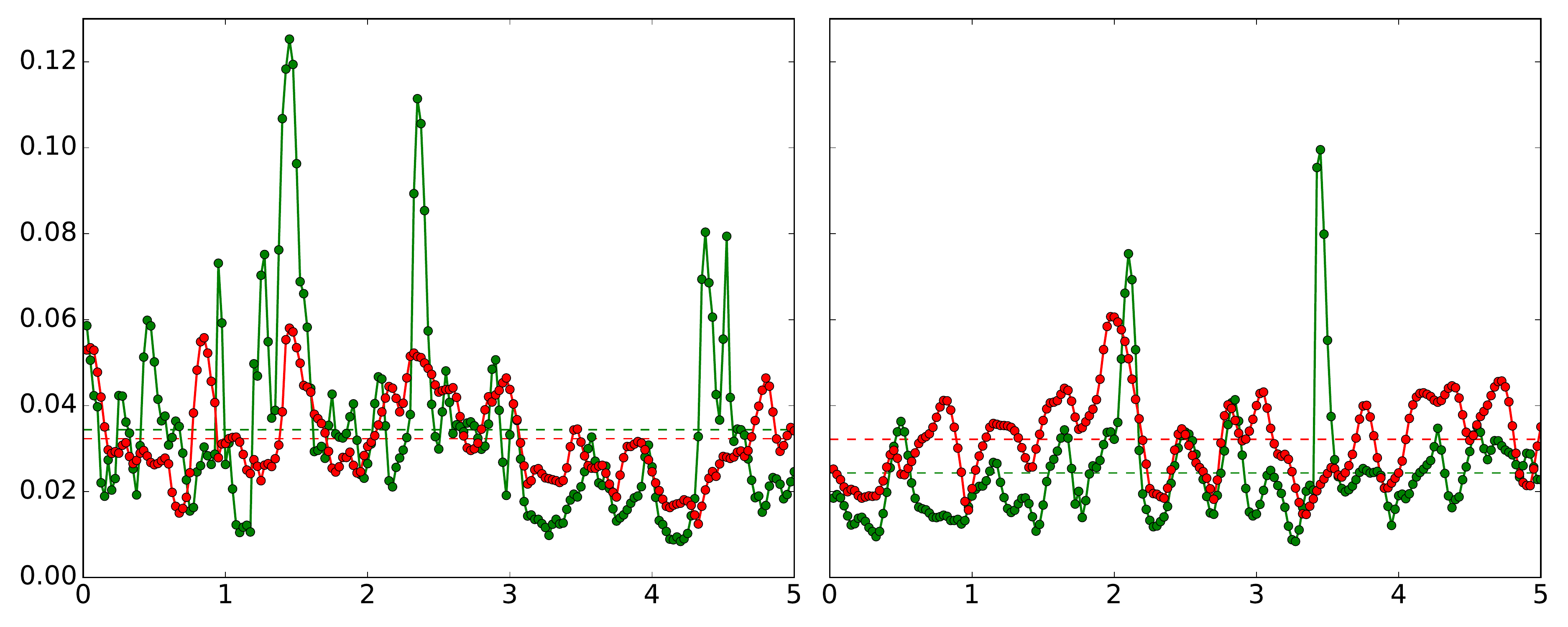}};    
\node (b) [below of=a, yshift=-2.5cm] {\includegraphics[trim={0cm 0.5cm 0cm 0},clip,width=1\linewidth]{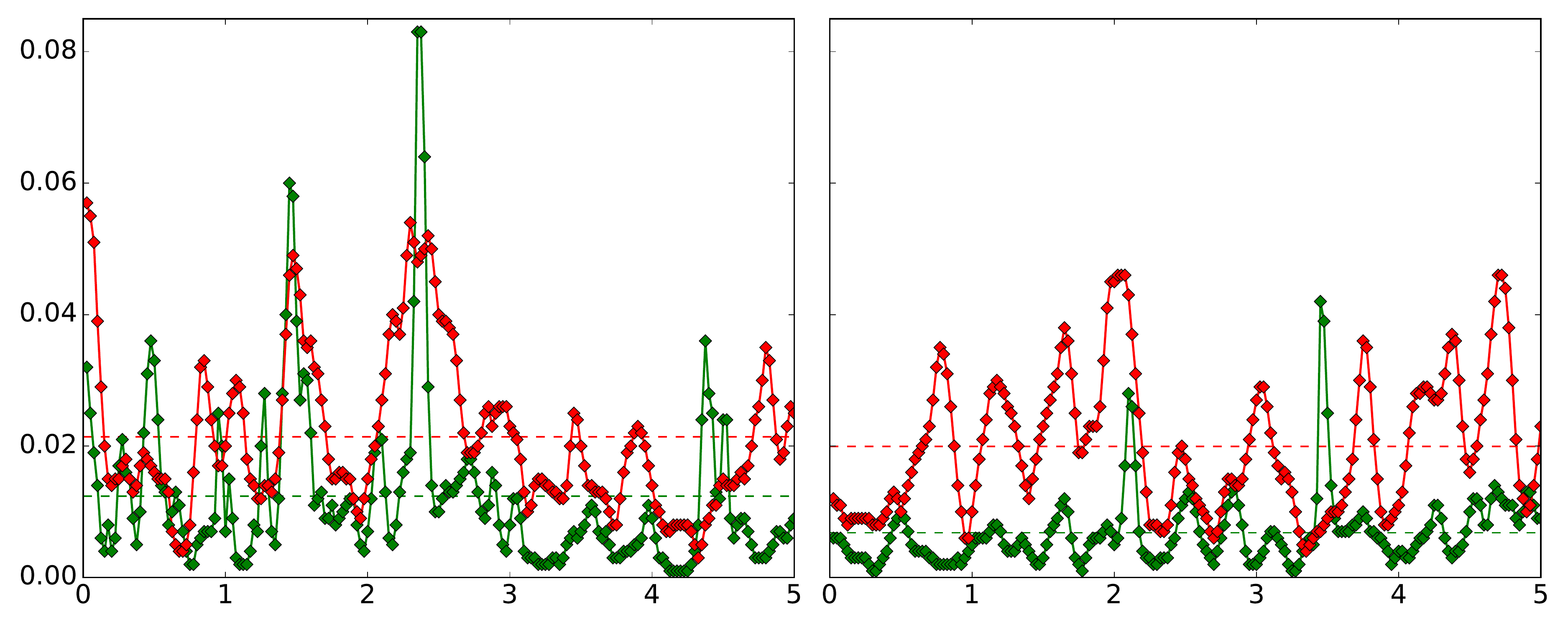}};    
\node (c) [xshift=0.21cm,yshift=-2.6cm] {\includegraphics[trim={0.95cm 24.0cm 14.25cm 0.95cm},clip,width=0.225\linewidth]{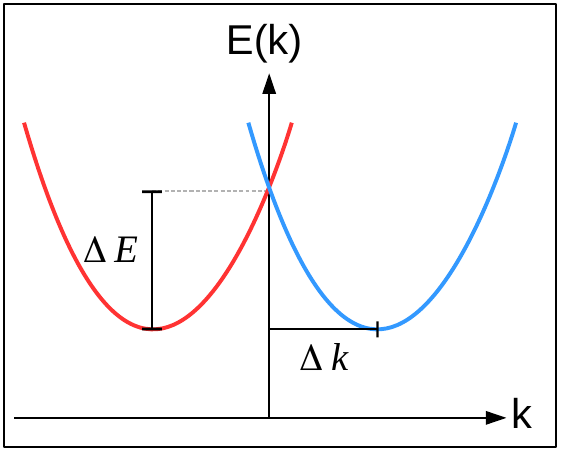}};
\node (d) [xshift=-0.8cm,yshift=1.0cm] {\includegraphics[trim={2.2cm 1.15cm 2.0cm 1.75cm},clip,width=0.165\linewidth]{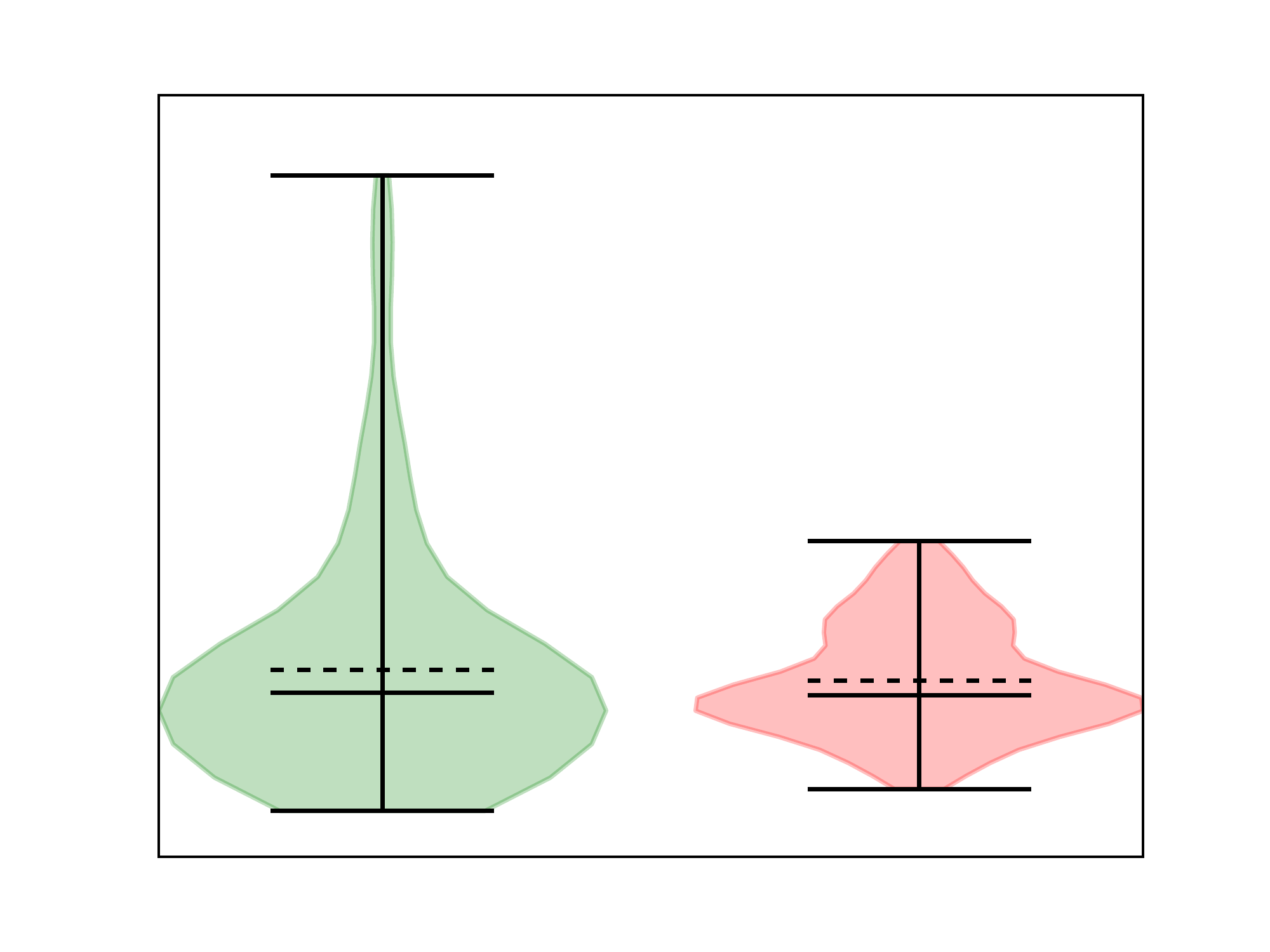}};
\node (e) [xshift=1.08cm,yshift=1.0cm] {\includegraphics[trim={2.2cm 1.15cm 2.0cm 1.75cm},clip,width=0.165\linewidth]{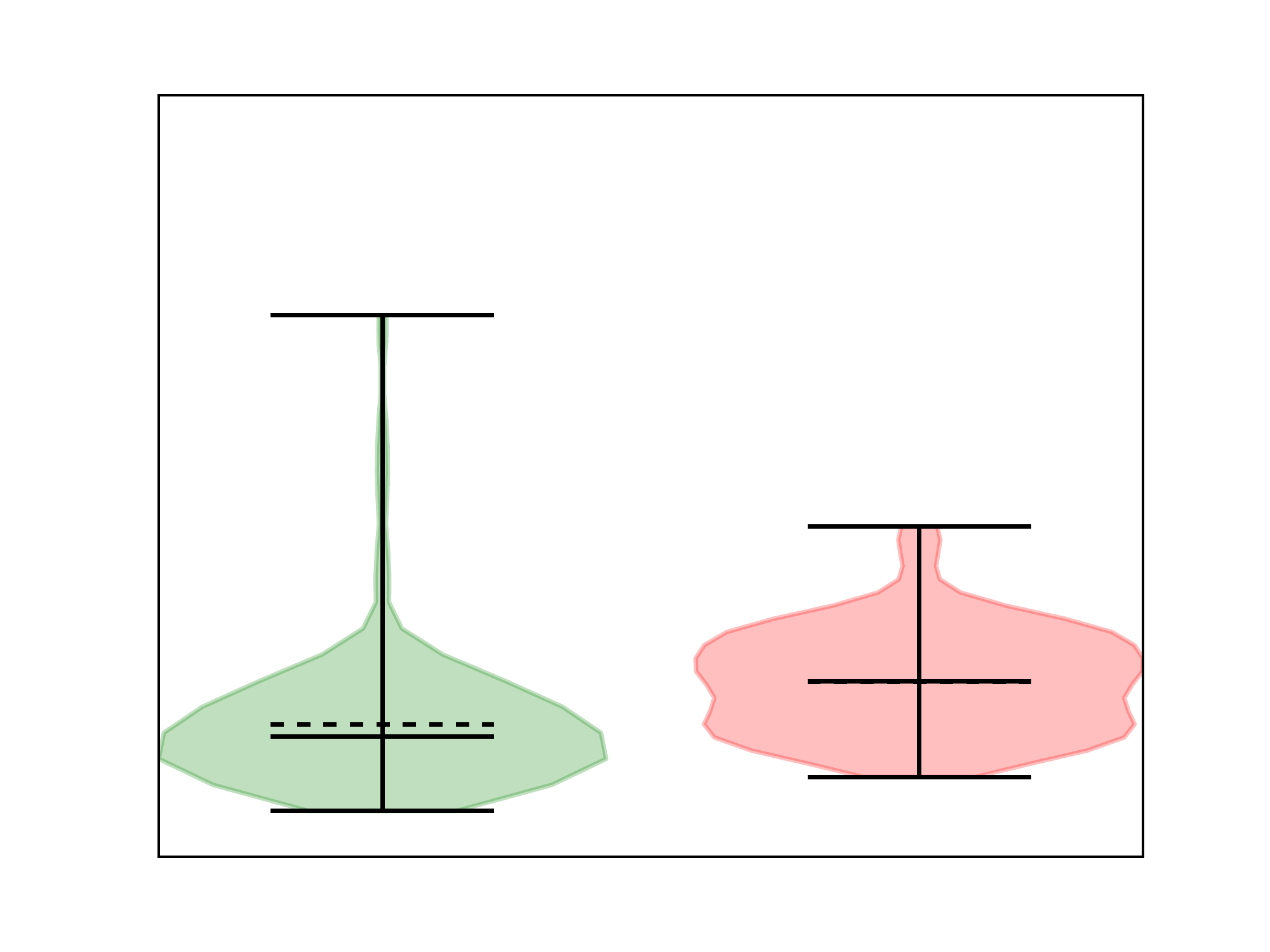}};
\node [above of=a, yshift=0.75cm, xshift=-1.9cm, rotate=0] {\textbf{a}};
\node [above of=a, yshift=0.75cm, xshift=+2.2cm, rotate=0] {\textbf{b}};
\node [left of=a, xshift=-3.5cm, rotate=90] {$\Delta k$ (\AA$^{-1}$)};
\node [left of=b, xshift=-3.5cm, rotate=90] {$\Delta$E (eV)};
\node [below of=b, xshift=-1.75cm, yshift=-1.0cm, rotate=0] {Time (ps)};
\node [below of=b, xshift= 2.3cm, yshift=-1.0cm, rotate=0] {Time (ps)};
\end{tikzpicture}
\caption{Magnitude of momentum and energy offsets in the valence and conduction bands for 
\textbf{(a)} \ce{CH3NH3PbI3} and 
\textbf{(b)} CsPbI$_{3}$ from 200 frames of \textit{ab initio} molecular dynamics. The valence band splittings
are plotted in green and the conduction band splitting in red.
The average splitting is plotted as dashed lines.
The splitting schematic, using conduction band as example, highlights the spin splitting parameters $\Delta E$ and $\Delta k$.
The probability densities of the momentum offsets are represented by inset violin plots, where the mean and median values 
are shown as dashed and solid lines. 
}
\label{f2}
\end{figure}

\begin{figure}
\centering
\includegraphics[]{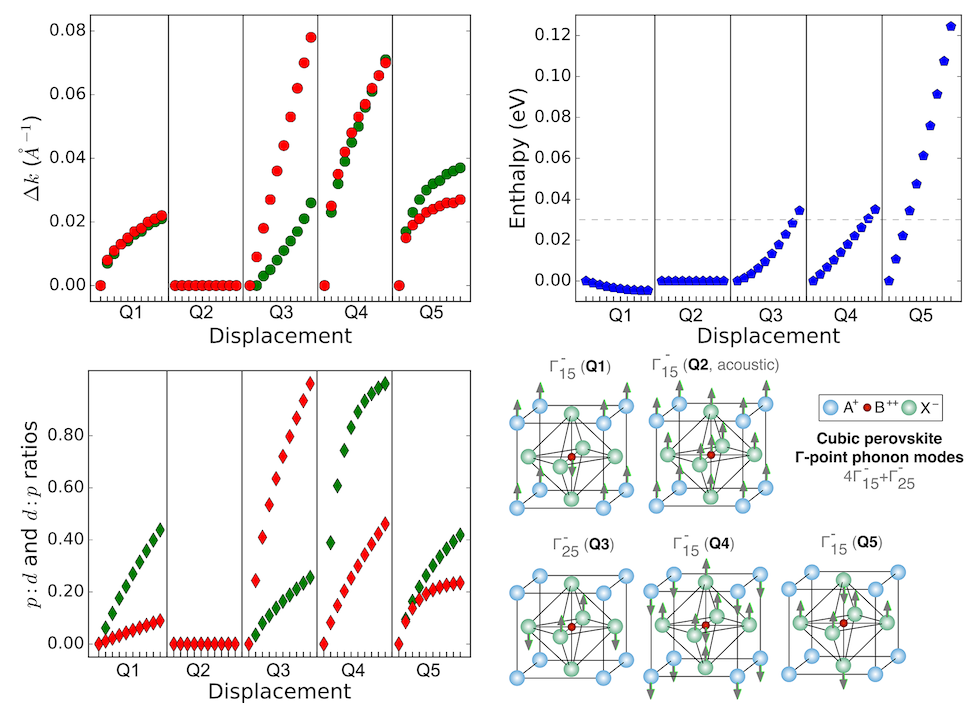}
\caption{Analysis of the five zone-centre phonon modes for cubic CsPbI$_{3}$.
The effect of displacement along each mode (Q) is shown for:
momentum offsets ($\Delta k$) in the valence (green) and conduction (red) bands;
enthalpy change relative to reference cubic structure ($k_B$T at 300 K is plotted as a dashed grey line);
ratio of $d:p$-character (green) in the valence band maximum and $p:d$-character (red) in the conduction band minimum 
eigenstates versus displacement. The ratios are computed from Mulliken populations and both sets are normalised for comparison.  
}
\label{f3}
\end{figure}

\textbf{Dynamic disorder at finite temperatures.}
Electronic structure calculations are typically athermal, the nuclear
coordinates approximated as classical, definite, points in space. 
To consider finite temperature effects we sample structures from
a thermodynamic ensemble. 
Within \textit{ab initio} MD, the classical equations of motion for the nuclear coordinates are
integrated with the forces from ground-state DFT
calculations. 
The kinetic energy (temperature) of these ions is connected to a thermostat
(chosen to be 300 K). 

For the MD, we consider 2$\times$2$\times$2 supercell expansions of the 
\ce{MAPbI3} and \ce{CsPbI3} unit cells, allowing for Brillouin zone boundary tilting modes.
After an initial equilibration period, 
we sample 200 consecutive frames, each separated by 0.025 ps. 
This time discretisation is intentionally chosen to be less than the
characteristic motion of the material (most particularly the 2 ps
methlyammonium rotation, and associated cage distortion). 
We can track the electronic structure and spin splitting evolve in time. 
For these snapshot structures, we calculate the electronic structure at the 
LDA level (all-electron with spin-orbit coupling). 
As discussed, we expect the momentum offset $\Delta k$ to be in close agreement with 
QS\textit{GW} for the CB, but largely overestimated in the VB. 

For both materials the momentum offset $\Delta k$ in the conduction band
(Fig. \ref{f2}) fluctuates around a value of 0.03 \AA$^{-1}$. 
A similar picture is seen 
for the energy offsets, with averages around 20 meV. In the valence band, larger 
differences in the average splittings are seen when comparing the two materials. The MA cation 
results in more variable behaviour, particularly in the VB, as evidenced by 
the larger spread (and broader distribution) in the inset violin plots. This is in line with the 
dominant contributions to the valence band coming from the lighter iodine atoms. 
Interestingly, the Cs cation, despite having the same average, results in a broader distribution 
in the conduction band.  
The splittings are reduced relative to the single cells but, crucially, are non-zero 
throughout.

As well as $\Delta k$ changing in magnitude, the band extrema rotate relative to one another.  
The average angle between the line connecting the pair of minima in the conduction 
band and the line connecting the pair of maxima in the valence band is 
44$^{\circ}$ for \ce{MAPbI3} and 38$^{\circ}$ for CsPbI$_{3}$ (relative angles are as high as 90$^{\circ}$).
A plot of the angular dependence (Fig. S2),  a scatter plot of the extremal point positions (Fig. S3), 
and a discussion of the sensitivity of the results can be found in the 
Supplemental Material. 
%

The similar band splittings observed for both cesium and methylammonium suggest 
that the dominant contribution is thermal distortion of the inorganic cage. 
A Fourier transform of the time-series suggests the spin splitting has the 
same periodicity as low frequency phonon modes (Fig. S4).
These modes are distortions of the octahedral cage and associated rotations of
the organic cation in the case of \ce{MAPbI3}.
As such, spin splitting is expected to be present at room temperature for all lead iodide
based perovskites, largely independent of cation or (long-range) crystal symmetry.

CsPbI$_{3}$ is a striking example of the importance of 
finite temperature lattice dynamics on the electronic structure. 
A static calculation predicts no spin splitting. 
Such a cubic unit cell agrees with 
the average structure determined from X-ray diffraction \cite{MOLLER1958};
however, this phase has a number of vibrational instabilities that give 
rise to dynamic distortions \cite{Yang2017a}.
At any point in time, the lattice 
distortions break the local symmetry, give rise to local fields, and produce
spin splitting.
To better understand the structural origins of the splitting, 
we return to a (single) unit cell representation and follow symmetry breaking by phonon modes. 
In the frozen-phonon approximation, we separately distort along each of the $\Gamma$-point normal modes
and track $\Delta k$. 
The magnitude of these distortions are chosen 
(from the phonon frequency) to sample a
thermodynamic degree of freedom to an energy of $k_B T$ (300 K).

The five unique cubic-perovskite phonon modes are depicted in Fig. 3. 
The acoustic mode (Q2), is a rigid translation of the 
lattice and does not change the energy or electronic structure;
it is included as a cross-validation. 
Each other mode results in a monotonic spin splitting as a function of displacement. 
The first mode is soft, corresponding to `rattling' of the
Cs and Pb ions. 
The dependence of $\Delta k$ on displacement has the same form for both VB and CB. 
Notable differences in $\Delta k$ are observed in both bands for Q3 and Q5.
The largest initial (low displacement, low energy) splitting is observed for
Q4, which corresponds to the Pb atom moving off site. 
Mode Q3 shows remarkably linear splitting as a function of distortion.  
The mode is composed of octahedral twisting (iodine motion) 
around Pb. 
There are considerably larger splittings in the conduction band than in 
the valence band.
The highest energy modes result in the largest splittings over the thermally accessible
range.

Spin-orbit coupling is commonly considered a local effect, significant around heavy atoms.  
Here, the conduction band is primarily Pb-$p$ orbital character and the valence band 
is primarily I-$p$ character. 
Spin-orbit coupling is large close to the Pb (Z=82) and 
I (Z=53) nuclei. The spin splitting has its origins in spin-orbit coupling, which appears 
in the non-relativistic limit of the Dirac equation. The contribution to the electronic 
energy levels is the expectation value of the following spin-orbit correction 
to the scalar relativistic Hamiltonian: 
$$H_{SOC}=\frac{2}{c^{2}}(\nabla V \times \mathbf{p})\cdot \mathbf{s}$$
where $V$ is the crystal potential, $\mathbf{p}$ is a momentum operator, $\mathbf{s}$ is a spin 
operator and $c$ is the speed of light. 
Spin splitting is a result of local asymmetry in both the potential 
 and the wavefunction (through the momentum operator)
in the core regions of heavy nuclei; a large splitting requires a large value for both 
\cite{Bihlmayer2006}. 

Previous work has shown that the asymmetry of the wavefunction strongly influences the 
magnitude of the spin splitting \cite{Bihlmayer2006,Nagano2009,Henk2010}. This can be 
estimated by the ratio of the $l$ to $l \pm 1$ angular-momentum character in a 
region close to the nuclear core. The Mulliken populations in Fig. 3 do not project out small 
regions close to the nuclei; nevertheless, the ratios correlate closely to the observed shifts in $\Delta k$. 
There is a dependence on the choice of $l$- to $l \pm 1$-character, with some trends, for example, better reproduced 
by the ratio of $p:s$ (see Fig. S5). 
The spin splitting is also found to depend on local electric fields (see Fig. S6). 
These are spherical averages around the Pb and  I locations (for iodine a vector average is taken). 
The angle between the electric field and the momentum offset vectors is
approximately 90$ \pm2^{o}$ in each case, in line with the perpendicular relation between the electric
field and momentum vectors in the Rashba model.  
The local symmetry breaking, resulting in asymmetry of both the potential and wavefunction, can be seen 
through changes in the charge density (contour plots are provided in Fig. S7). 

To quantify the correlation for disordered structures, 
we calculate these measures (at the DFT level) for 147 random methylammonium orientations 
in a pseudocubic unit cell. 
The MA orientation is randomly sampled and the \ce{PbI3} network is then locally
relaxed around each configuration. 
Optimized structures with energy $> k_BT$ were excluded. 
The $d:p$ and $p:d$ ratios are found to correlate well with the magnitude of the 
momentum offset $\Delta k$ (see Fig. 4).  
Improved agreement is expected by accounting for different 
$l$ to $l \pm 1$ ratios 
\footnote{For instance, Pearson correlation coefficient values are as high as 0.94 in the valence band 
for $p:s$ and $p:d$ ratios based on projections for iodine alone.} 
and by limiting the projections to regions close to the nuclear cores. 

In addition to band splitting, structural distortions generally widen the bandgap and flatten
the frontier bands.  QS\textit{GW} \emph{underestimates} the bandgap
in CsPbI$_{3}$ in the cubic structure, while almost universally
it slightly \emph{overestimates} gaps in semiconductors and
insulators \cite{van2006quasiparticle}.  However, QS\textit{GW}
predicts an increase of 0.44\,eV in the pseudocubic
structure relative to the cubic one (Table 1), improving
agreement with room temperature experiments.  Average \ce{MAPbI3} and 
\ce{CsPbI3} bandgaps from MD simulations also increase by
0.25 and 0.73 eV relative to athermal structures, and they
increase for all displacements along the four CsPbI$_{3}$
optical phonon modes (See Fig. S8).
These trends originate from two main sources: (1) the antibonding
character of the frontier bands \cite{Huang2013} and (2) a
reduction in spin-orbit splitting of the CB
originating from a reduction in Pb-$p$ content.
Both trends are demonstrated through an analysis of the
shifts in the  band edges with
displacements along the normal modes.  
There is a general tendency
for the valence band to be pushed \emph{down} and for the
conduction band to be pushed \emph{up}, though how each edge
shifts depends on the normal mode (see Fig. S8).  
The downshift of the valence band is in line with a reduction in antibonding 
between I-$p$ and Pb-$s$ states \cite{Huang2013}. 
On the other hand, the CB upshift correlates closely with a reduction 
in Pb-$p$ character and is largely a result of the reduced SO coupling 
of the state (see Fig. S9).         


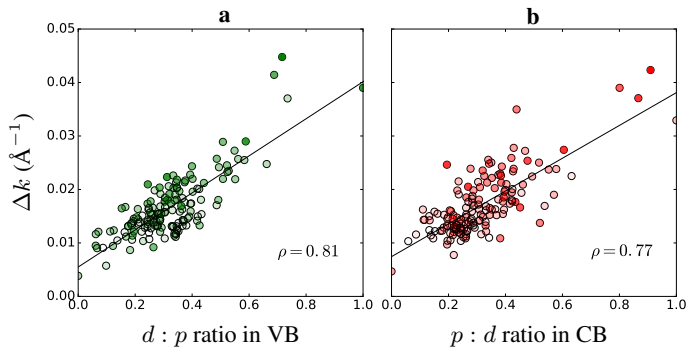
\begin{figure}
\centering
\begin{tikzpicture}
\node (a) {\includegraphics[trim={3cm 1cm 3cm 1cm},clip,width=1\linewidth]{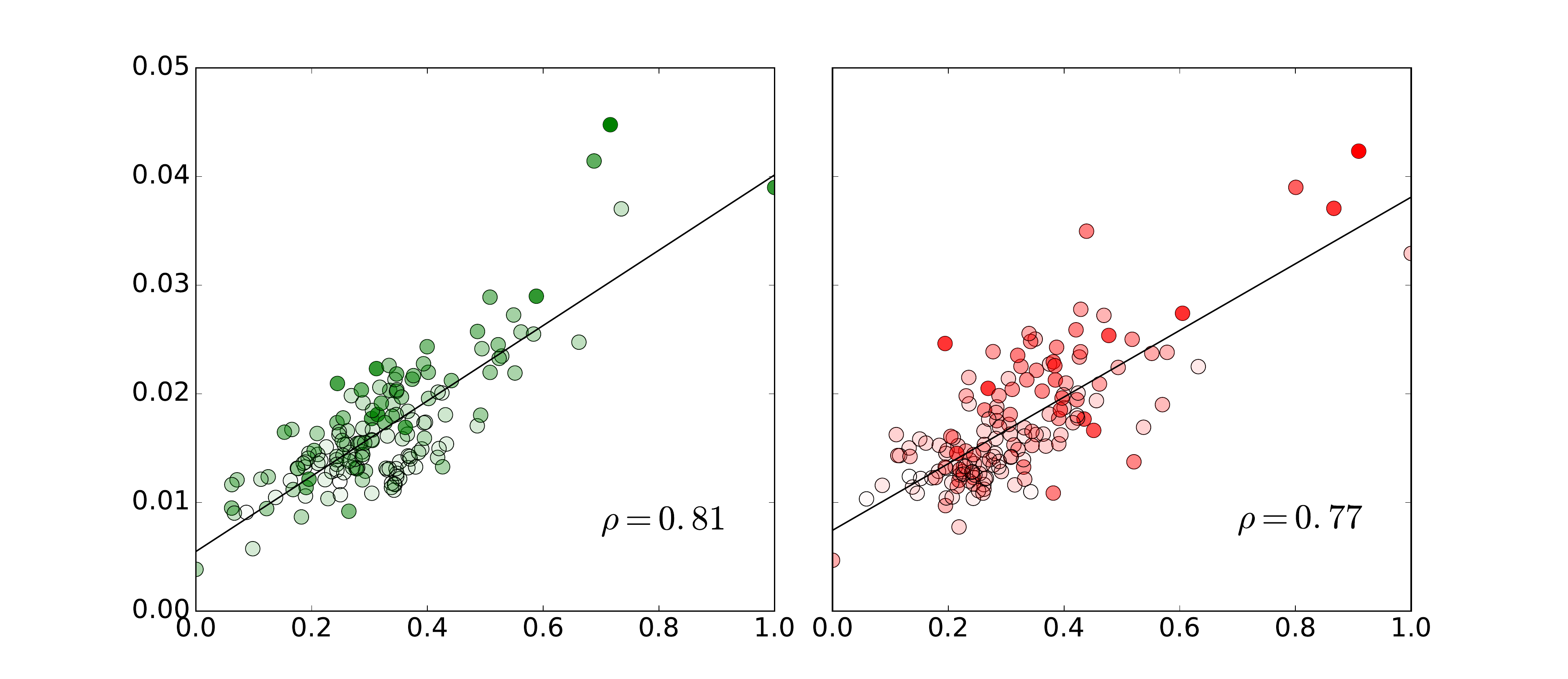}};  
\node [above of=a, yshift=0.95cm, xshift=-1.9cm, rotate=0] {\textbf{a}};
\node [above of=a, yshift=0.95cm, xshift=+2.2cm, rotate=0] {\textbf{b}};
\node [left of=a, xshift=-3.6cm, rotate=90] {$\Delta k$ (\AA $^{-1}$)};
\node [below of=a, xshift=-1.95cm, yshift=-1.3cm, rotate=0] {$d:p$ ratio in VB};
\node [below of=a, xshift=+2.15cm, yshift=-1.3cm, rotate=0] {$p:d$ ratio in CB};
\end{tikzpicture}
\caption{Magnitude of momentum offset $\Delta k$ versus \textbf{(a)} $d:p$ ratio in the VB and 
\textbf{(b)} $p:d$ ratio in the conduction band for 147 stochastically sampled
MA orientations in MAPbI$_{3}$. The total energy of each structure is represented by the opacity 
level of the marker face colour, where higher opacity refers to higher energy. The Pearson correlation 
coefficient ($\rho$) between the splitting and ratio magnitudes is shown in the bottom right
 corner of the plots.}
 \label{f4}
\end{figure}


In summary, a spin-split indirect gap has been found in thermal ensembles of
organic-inorganic and inorganic lead halide perovskites.
The splitting arises due to relativistic spin-orbit coupling in the presence 
of local asymmetry, which we relate to distortions in the lead iodide framework.  
The role of the perovskite A-site cation appears to be steric in nature. 
By studying the effect of distortions along individual phonon modes, we uncover
a mostly linear relationship between the energy of a distortion and the momentum offset.
The asymmetry in the electronic wavefunction and, to a lesser extent, the local electric field were 
found to correlate well with the magnitude of the momentum offset.
Overall, structural distortions lead to spin-split bands and a widening of the bandgap. 
This provides a method to go from an observed (temperature-dependent)
disorder measure to a measure of the indirect gap. 
Temperature-dependent observations of the band character for both the organic
and inorganic perovskites will provide information to confirm these hypotheses. 

~

In addition to the data in the Supplemental Material, 
a collection of input and output files, including
crystal structures, are available in an on-line repository [DOI inserted
upon acceptance].

~

Support has been received from the EPSRC (projects EP/M009580/1, EP/M011631/1 and EP/K016288/1).
Via our membership of the UK's HEC Materials Chemistry Consortium, which is funded by EPSRC (EP/L000202), this work used the ARCHER UK National Supercomputing
Service (http://www.archer.ac.uk).

\bibliography{MAPI-local-field}

\onecolumngrid 
\clearpage

        \setcounter{table}{0}
        \renewcommand{\thetable}{S\arabic{table}}%
        \setcounter{figure}{0}
        \renewcommand{\thefigure}{S\arabic{figure}}%
        \setcounter{equation}{0}
        \renewcommand{\theequation}{S\arabic{equation}}%

{\LARGE{Supplemental Material for `Dynamic Symmetry Breaking and Spin Splitting
in Metal Halide Perovskites'}}

\section*{Additional Computational details}
Electronic structure calculations are carried out using the full-potential linearized muffin-tin orbital (FP-LMTO) method \cite{dreysse2000electronic,mark06adeq} 
as implemented in the QUESTAAL code \cite{questaal}. All results presented use either the local density approximation (LDA) \cite{kohn1965self} or the 
quasiparticle self-consistent G$W$ (QSGW) method \cite{van2006quasiparticle}, using the LDA as the starting point. The basis set consists of two 
smoothed Hankel envelope functions per $l$ channel and local orbitals are included for Pb (5$d$), Cs (5$s$) and 
I (high-lying $s$ and $p$). Additional `floating orbitals' up to $l=1$ are also included to provide a better description of the 
interstitial regions. $6 \times 6 \times 6$ and $4 \times 4 \times 4$ $k$-point meshes are used for single cell and 
supercell DFT calculations respectively. A $4 \times 4 \times 4$ $k$-point mesh is used in the QSG$W$ calculations. 
The QSG$W$ bandgaps were found to be sensitive to the product basis tolerance and more conservative settings were used. 
Free atom calculations (see Table S1) are carried out using both scalar Dirac and fully relativistic Dirac formulations  \cite{Ebert89}. 

A key finding in our work is that the magnitude of the spin-split momentum
offset is sensitive to the structure.  
Our ab-inito molecular dynamics is conducted with the \textsc{Vasp} codes. 
This is a plane-wave code that makes use of pseudo-potentials. 
A key necessity is to use pseudopotentials (i.e. {\tt Pb\_d}) which treat the
d-electrons explicitly, in order to get a correct representation of the dynamic
bonding environment and crystal field of the lead ion.

In our prior work
\cite{Azarhoosh2016}, we briefly studied samples from
ab-initio molecular dynamics on 2$\times$2$\times$2 unit cell with
$\Gamma$-point sampling and a small basis set (300 eV cutoff). 
It is generally expected that using such underconserved values in a plane wave
code has the effect of softening the potential energy landscape, leading to an
artificially  higher effective temperature (greater thermal fluctuations). 
After tests, we found that the lead thermal displacements were uniquely
sensitive to the reciprocal space sampling. 
Using a 3$\times$3$\times$3 $k$-point integration, we find reduced structural
distortions and splittings, compared to our earlier work.

Representative single-unit cells with randomised methylammonium orientation
were generated in a stochastic manner. 
This followed a prescription we had developed for and applied to stochastic
sampling of the phonon spectrum\cite{Leguy2016}. 
Starting with an optimised pseudo-cubic
structure, the methylammonium was removed, and then randomly reorientated by
applying a rotation matrix generated from a quarternion populated with
pseudo-random normally distributed values. A normally distributed shift
(standard deviation of 0.05 lattice vectors) was also applied to the
methylammonium to further randomise the starting configuration.  101 steps of
conjugate gradient optimisation of the ionic locations were then applied to
these structures, with the PBESol functional and a 700 eV cutoff in the kinetic
energy. This local optimisation procedure first fixes the `bad contacts'
between overlapping atoms, but due to the local nature ends up in a local
representative minima. The resulting structures were extracted, along with the
final DFT total energy. This produced 249 unique structures. 

The DFT total energies were then assessed. A partition function was
calculated by summing over the sampled DFT energies. Examples with a
thermodynamic probability of p $<1$e$^{-5}$ compared to the lowest energy structure
were discarded. At our chosen temperature T = 300 K, this is an energy cutoff of 297.6 meV.
The result was an ensemble of 147 athermal structures, representative of a
thermodynamic ensemble at 300 K.

\begin{table}[h]
\begin{center}
 \begin{tabular}{| c | c c c c c c|}
 \hline
 \multicolumn{1}{|c|}{} & \multicolumn{6}{c|}{Atomic energy levels (eV)} \\                       
                             & Pb   &   I  &  N  &  C  &  H  &  \\
 \hline
 $s$                           &  -12.4 &    -17.73  &  -18.5  &  -13.73  &  -6.44  &  \\
 $s_{\frac{1}{2}}$           &  -12.31 &    -17.68  &  -18.5  &  -13.72  &  -6.44  &  \\
 $p$                           &  -3.79 &    -7.33  &  -7.32  &  -5.5  &    &  \\
 $p_{\frac{1}{2}}$           &  -4.87 &    -8.0  &  -7.33  &  -5.51  &    &  \\
 $p_{\frac{3}{2}}$           &  -3.36 &    -6.97  &  -7.32  &  -5.5  &    &  \\
 $p - p_{\frac{1}{2}}$     &  1.08 &    0.68  &  0.01  &  0.0  &    &  \\
 \hline
\end{tabular}
\end{center}
\caption{Relativistic atomic energy levels. The $s$ and $p$ refer to the standard scalar-relativistic energy levels (without SOC), while $s_{\frac{1}{2}}$, $p_{\frac{1}{2}}$ 
and $p_{\frac{3}{2}}$ are the fully-relativistic Dirac energy levels. The $p - p_{\frac{1}{2}}$ is a measure of the gap reduction as a result of relativistic effects.}
\end{table}

\begin{table}
\begin{center}
 \begin{tabular}{| c | c c c|}
 \hline
 \multicolumn{1}{|c|}{} & \multicolumn{3}{c|}{Effective masses} \\ 
 \multicolumn{1}{|c|}{} & \multicolumn{1}{c}{VB} & \multicolumn{2}{c|}{CB} \\ 
 \hline
                  &                    & &                     \\
 \ce{MAPbI3} orthorhombic          & -0.17,   -0.21,   -0.22  & &  0.22,     0.21,    0.13   \\
 \ce{MAPbI3} tetragonal            & -0.81,   -0.86,   -2.82  & &  26.17,    0.96,    0.55 \\
 \ce{MAPbI3} pseudocubic           & -0.16,   -0.31,   -0.50  & &  0.97,     0.17,    0.14 \\
 \ce{FAPbI3} pseudocubic           & -0.11,   -0.17,   -0.23  & &  0.15,     0.14,    0.12 \\
 \ce{CsPbI3} cubic                 & -0.12,   -0.12,   -0.12  & &  0.12,     0.12,    0.12 \\
 \ce{CsPbI3} pseudocubic           & -0.15,   -0.16,   -0.79  & &  0.60,     0.18,    0.14 \\ 
 \hline 
                  &                    & &                     \\ 
 Mode 1           &  -0.03, -0.04, -2.54  & &   2.03, 0.03, 0.03  \\
 Mode 2           &  -0.02, -0.02, -0.02  & &   0.02, 0.02, 0.02  \\
 Mode 3           &  -0.16, -0.25, -0.43  & &   0.18, 0.12, 0.12  \\
 Mode 4           &  -0.10, -0.17, -0.98  & &   0.36, 0.11, 0.08  \\
 Mode 5           &  -0.07, -0.11, -1.26  & &   1.32, 0.12, 0.06  \\
 \hline
\end{tabular}
\end{center}
\caption{Valence and conduction band effective masses. The three components m$_{1}$, m$_{2}$, m$_{3}$ are 
the effective masses along the principal axes of an ellipsoidal fit. 
The first six rows show results from QSG$W$ calculations. 
The following five rows contain LDA effective masses for \ce{CsPbI3} cubic  following the maximum (thermal) displacement 
along each of the five phonon modes. 
Note that Mode 2 (acoustic) corresponds to an undistorted cubic cell.}
\end{table}

\begin{figure}
\centering
\begin{tikzpicture}
\node (a) {\includegraphics[trim={1cm 1.5cm 1cm 2cm},clip,width=0.3\linewidth]{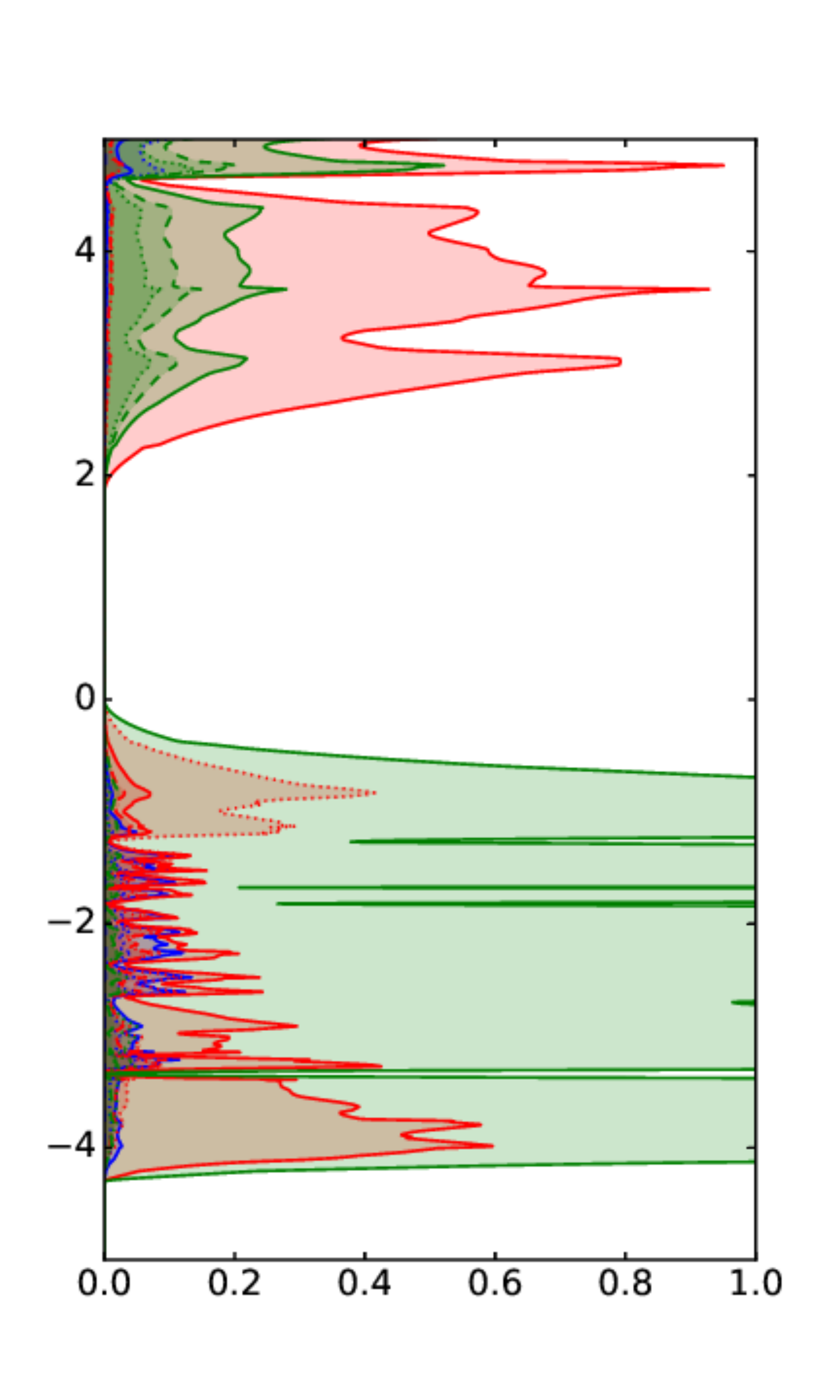}};
\node (b) [right of=a, yshift=0cm, xshift=4cm] {\includegraphics[trim={1cm 1.5cm 1cm 2cm},clip,width=0.3\linewidth]{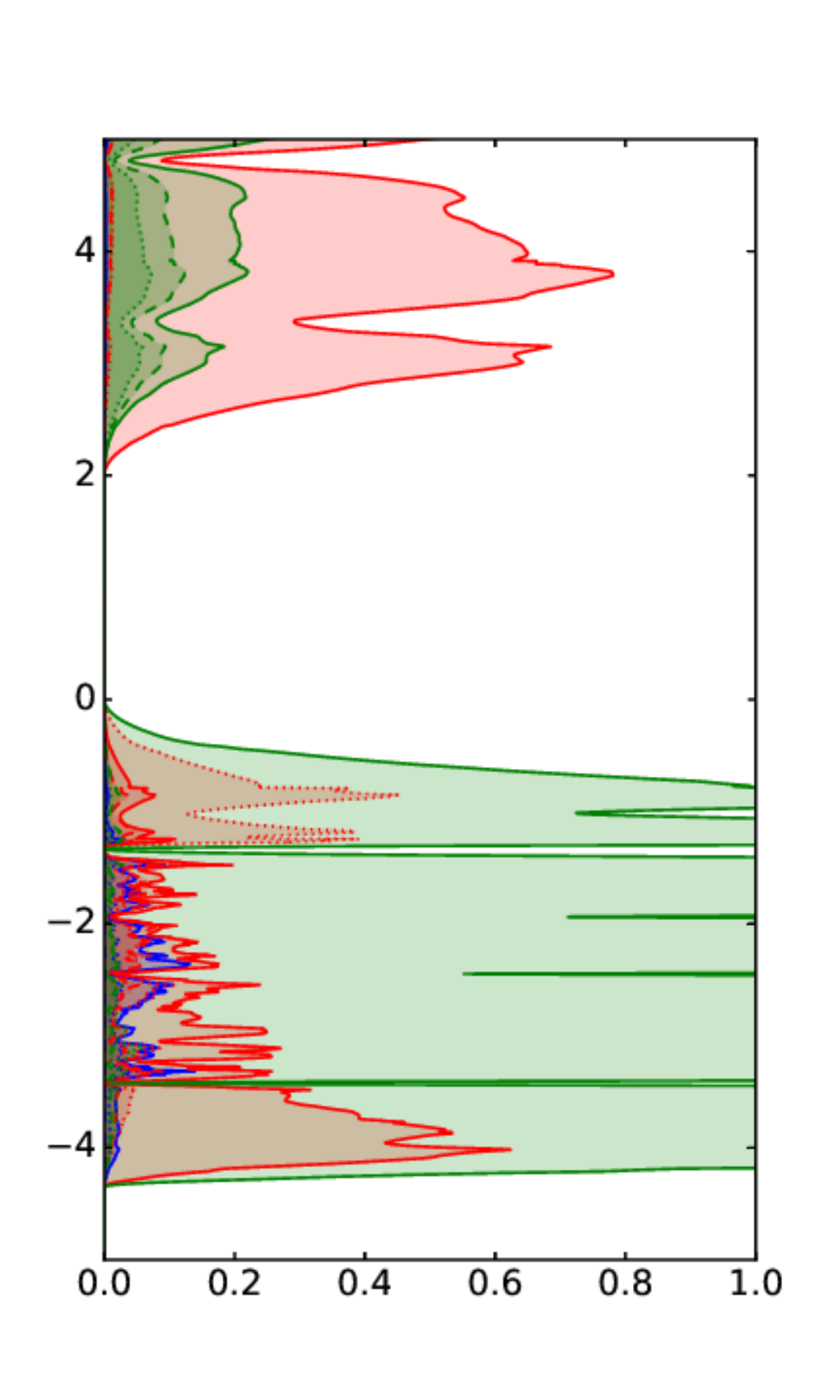}};
\node (c) [right of=b, yshift=0cm, xshift=4cm] {\includegraphics[trim={1cm 1.5cm 1cm 2cm},clip,width=0.3\linewidth]{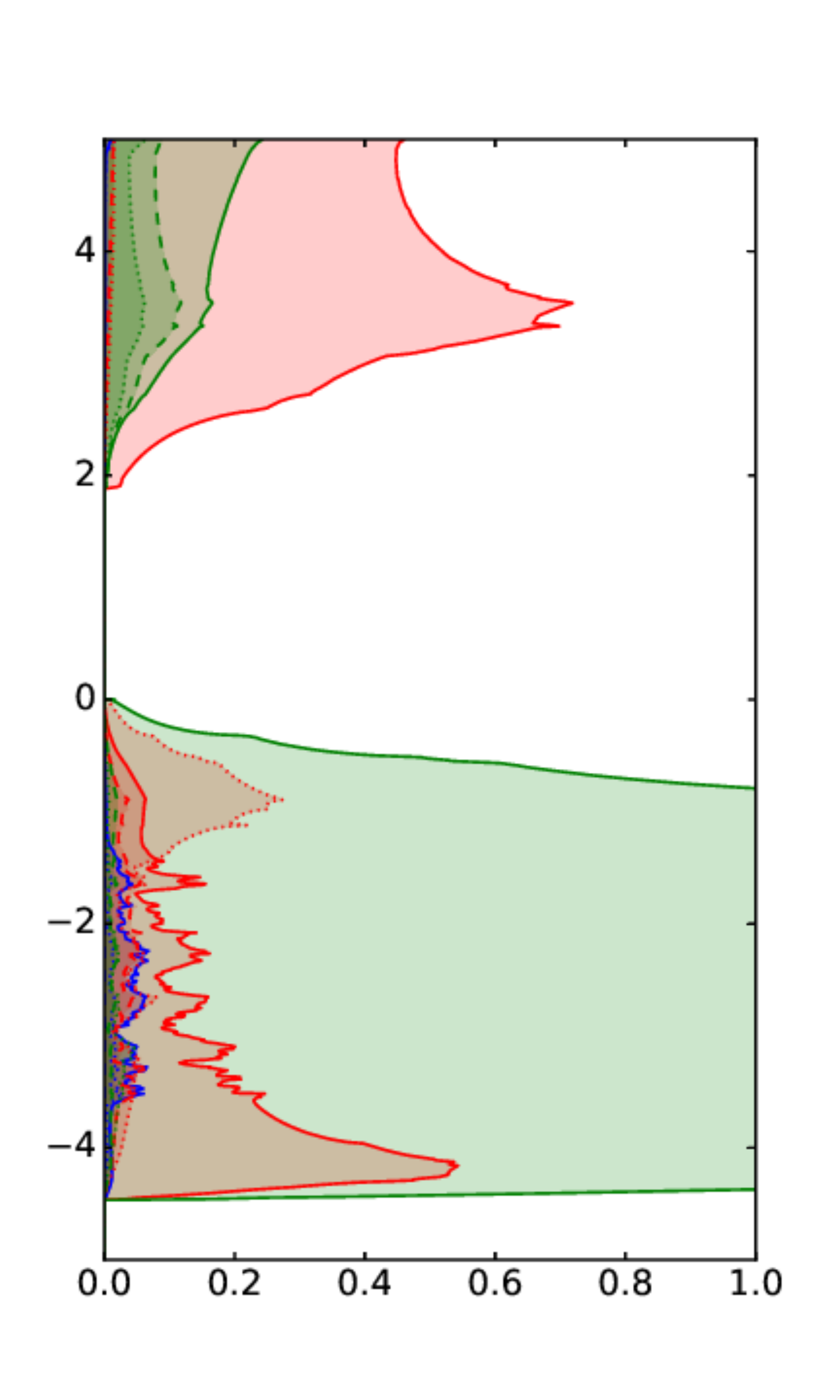}};
\node (d) [below of=a, yshift=-6.80cm, xshift=0cm]     {\includegraphics[trim={1cm 1.5cm 1cm 2cm},clip,width=0.3\linewidth]{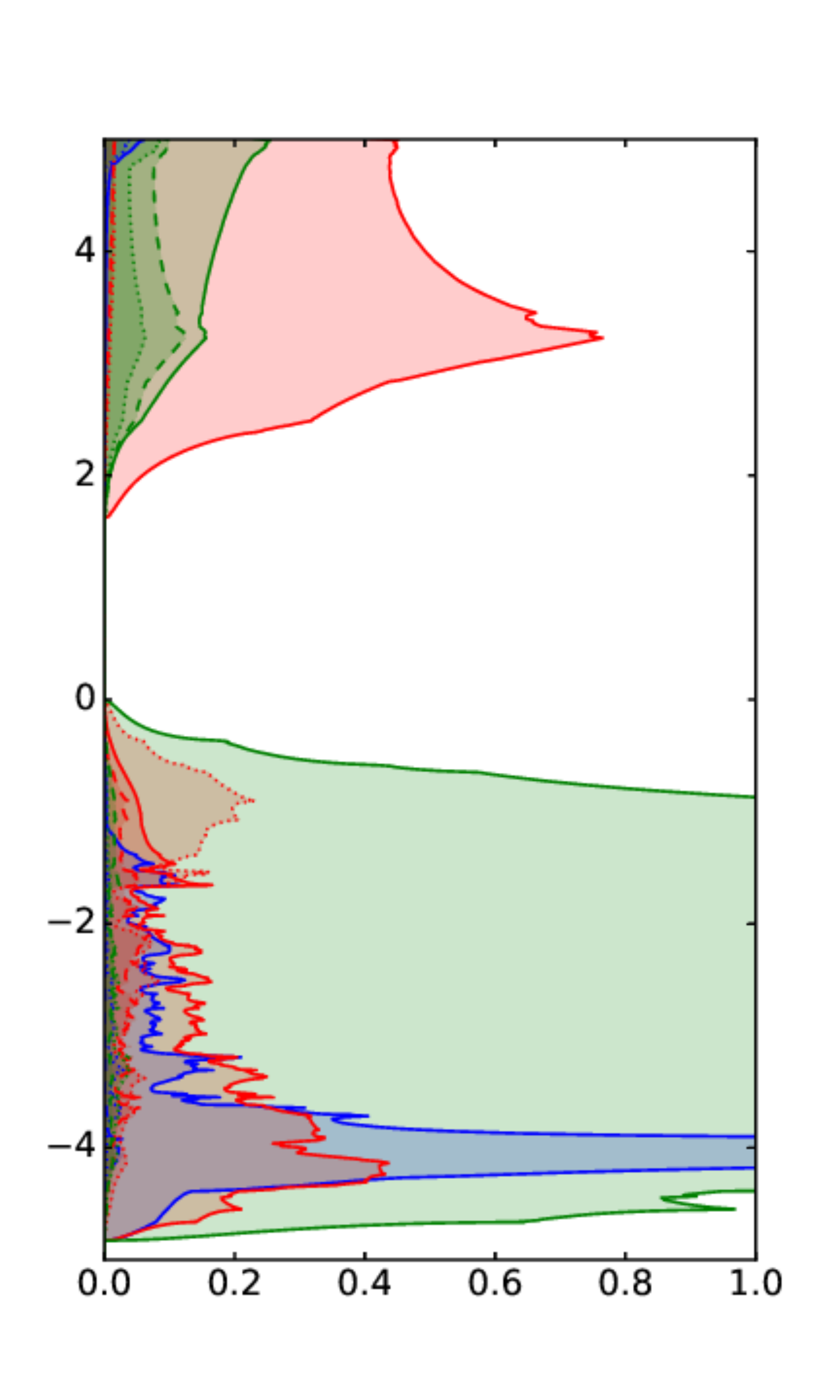}};
\node (e) [right of=d, yshift=0cm, xshift=4cm] {\includegraphics[trim={1cm 1.5cm 1cm 2cm},clip,width=0.3\linewidth]{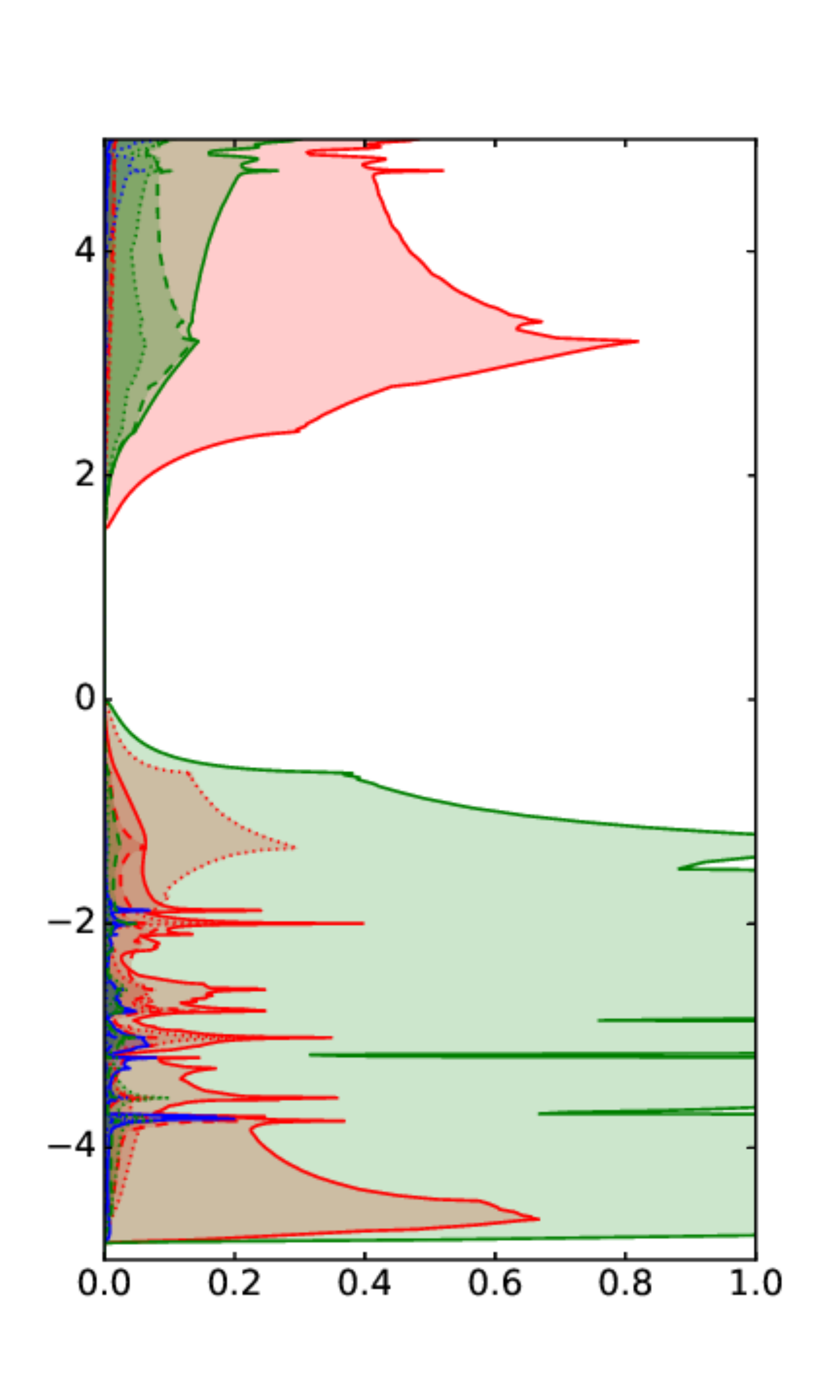}};
\node (f) [right of=e, yshift=0cm, xshift=4cm] {\includegraphics[trim={1cm 1.5cm 1cm 2cm},clip,width=0.3\linewidth]{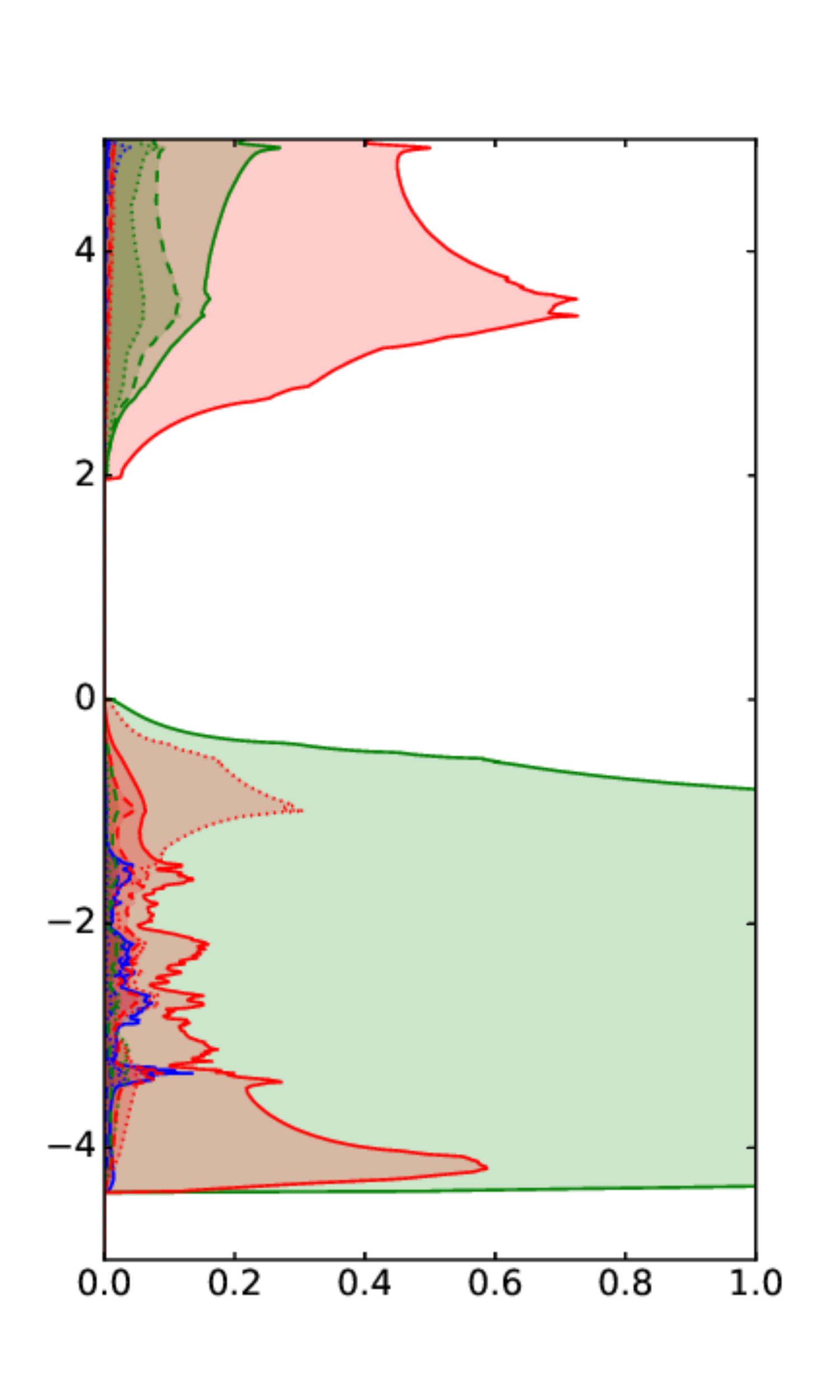}};
\node [above of=a, yshift=3.08cm] {\textbf{a}};
\node [above of=b, yshift=3.08cm] {\textbf{b}};
\node [above of=c, yshift=3.08cm] {\textbf{c}};
\node [above of=d, xshift=1.65cm,yshift=2.18cm] {\textbf{d}};
\node [above of=e, xshift=1.65cm,yshift=2.18cm] {\textbf{e}};
\node [above of=f, xshift=1.65cm,yshift=2.18cm] {\textbf{f}};
\node [left of=a, xshift=-1.65cm, rotate=90] {Energy (eV)};
\node [left of=d, xshift=-1.65cm, rotate=90] {Energy (eV)};
\node [below of=d, yshift=-3.25cm, rotate=0] {PDOS ($\text{eV}^{-1}$)};
\node [below of=e, yshift=-3.25cm, rotate=0] {PDOS ($\text{eV}^{-1}$)};
\node [below of=f, yshift=-3.25cm, rotate=0] {PDOS ($\text{eV}^{-1}$)};
\end{tikzpicture}
\caption{Partial electronic density of states from QS\textit{GW} calculations. The decomposition is based on a partial wave analysis (projection onto 
partial waves in augmentation spheres). Pb, I and cation (MA, FA or Cs) contributions are colored red, green and blue respectively. The 
$s$, $p$ and $d$ orbitals are denoted by dotted, solid and dashed lines. 
The first row, starting from the left, contains \ce{MAPbI3} in its (a) orthorhombic, 
(b) tetragonal and (c) pseudocubic phases.
The second row shows \ce{FAPbI3} in a (d) pseudocubic structure and CsPbI$_{3}$ in (e) cubic and (f) pseudocubic
structures. The pseudocubic CsPbI$_{3}$  is obtained by substituting Cs into \ce{MAPbI3}.}
\end{figure}

\begin{figure}
\centering
\begin{tikzpicture}
\node (a) {\includegraphics[trim={0cm 0.5cm 0cm 0},clip,width=1\linewidth]{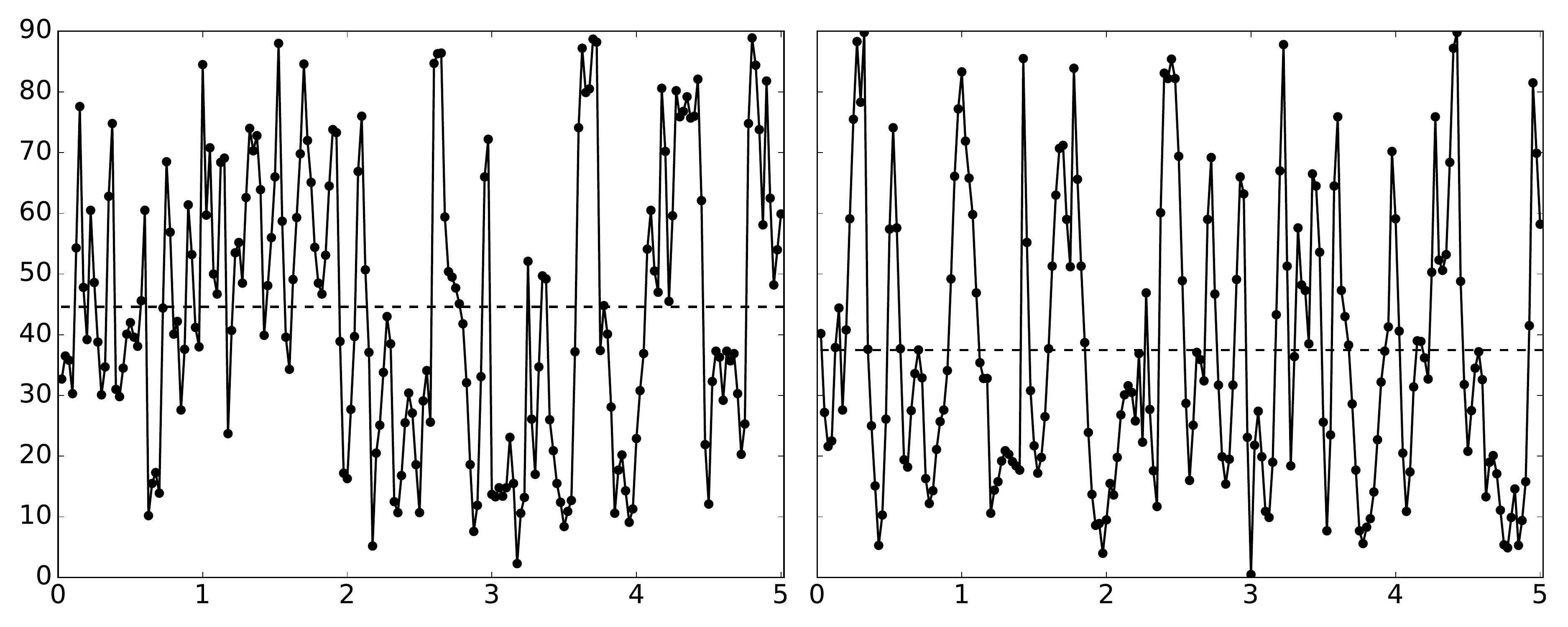}};    
\node [left of=a, xshift=-7.4cm, rotate=90] {$\theta$ (degrees)};
\node [below of=a, xshift=-3.2cm, yshift=-2.5cm, rotate=0] {Time (ps)};
\node [below of=a, xshift= 3.5cm, yshift=-2.5cm, rotate=0] {Time (ps)};
\end{tikzpicture}
\caption{Angle between momentum offsets in valence and conduction bands versus time. 
The angle is between the line connecting the pair of minima in 
the conduction band and line connecting the pair of maxima in the valence band. 
Average angles are shown as dashed lines.
}
\end{figure}

\begin{figure}
\centering
\begin{tikzpicture}
\node (a) {\includegraphics[trim={6cm 4cm 4cm 5cm},clip,width=0.45\linewidth]{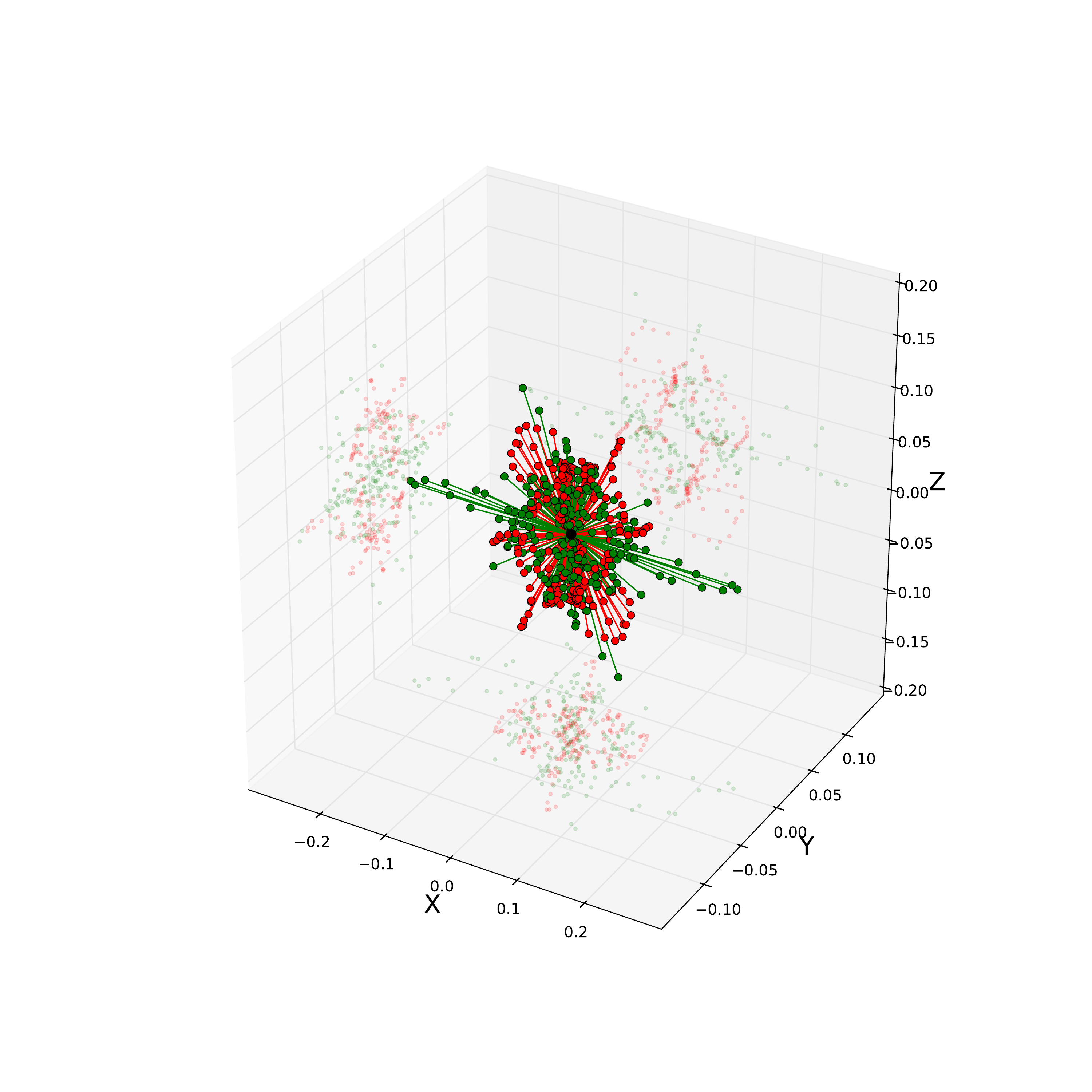}};
\node (b) [right of=a, yshift=0cm, xshift=6.5cm] {\includegraphics[trim={6cm 4cm 4cm 5cm},clip,width=0.45\linewidth]{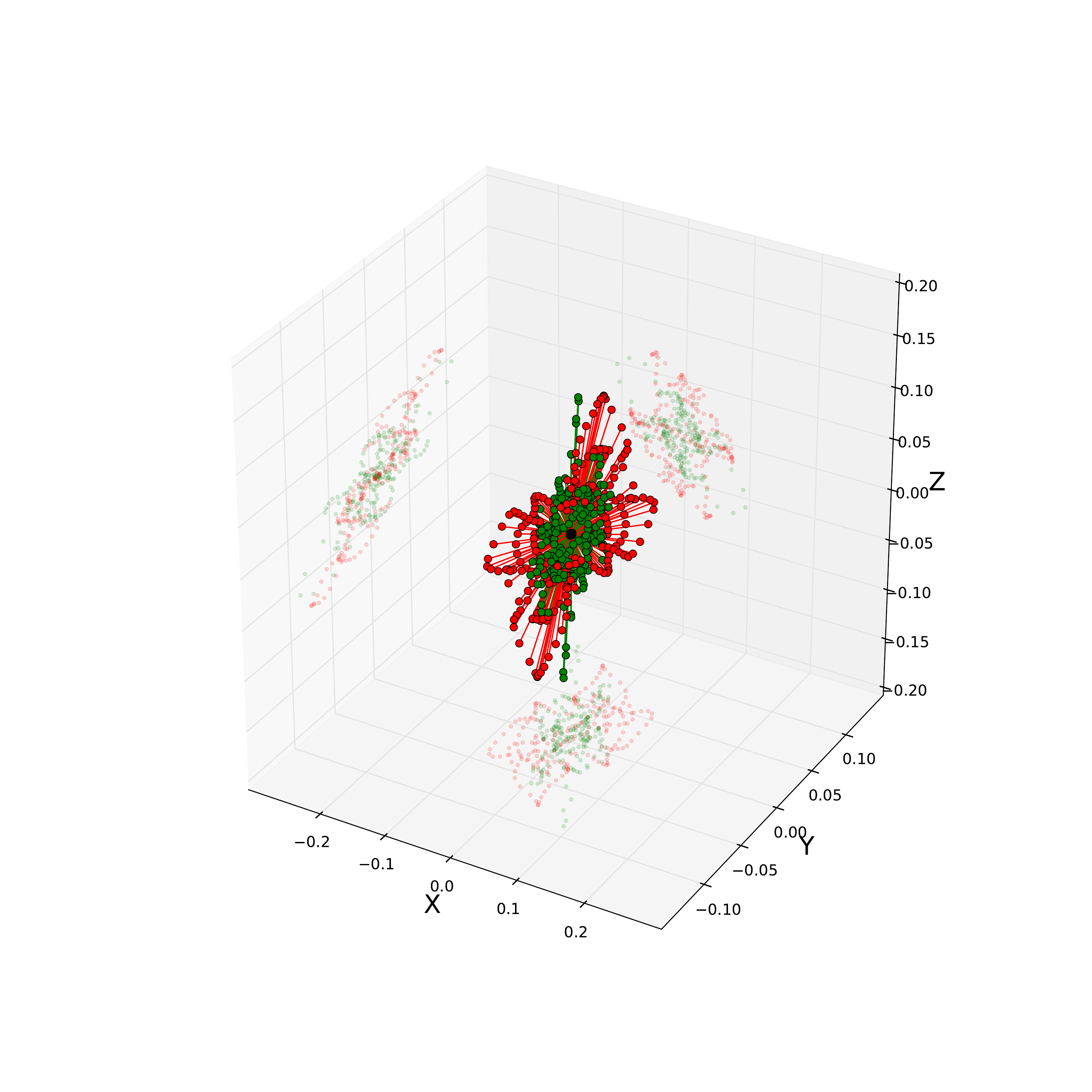}};
\end{tikzpicture}
\caption{Scatter plot of extremal point positions in \ce{MAPbI3} (left) and \ce{CsPbI3} (right) for 100 frames (first 2.5 ps of the 5 ps set). 
Four points and two lines are plotted from each frame: pair of maxima in the valence band (green), pair of minima in the conduction band (red) 
and lines connecting each extremal pair. Projections onto the $x$, $y$ and $z$ axes are also shown. 
}
\end{figure}

\begin{figure}
\centering
\begin{tikzpicture}
\node (a) {\includegraphics[trim={2.5cm 0.5cm 2.5cm 0.5cm},clip,width=1\linewidth]{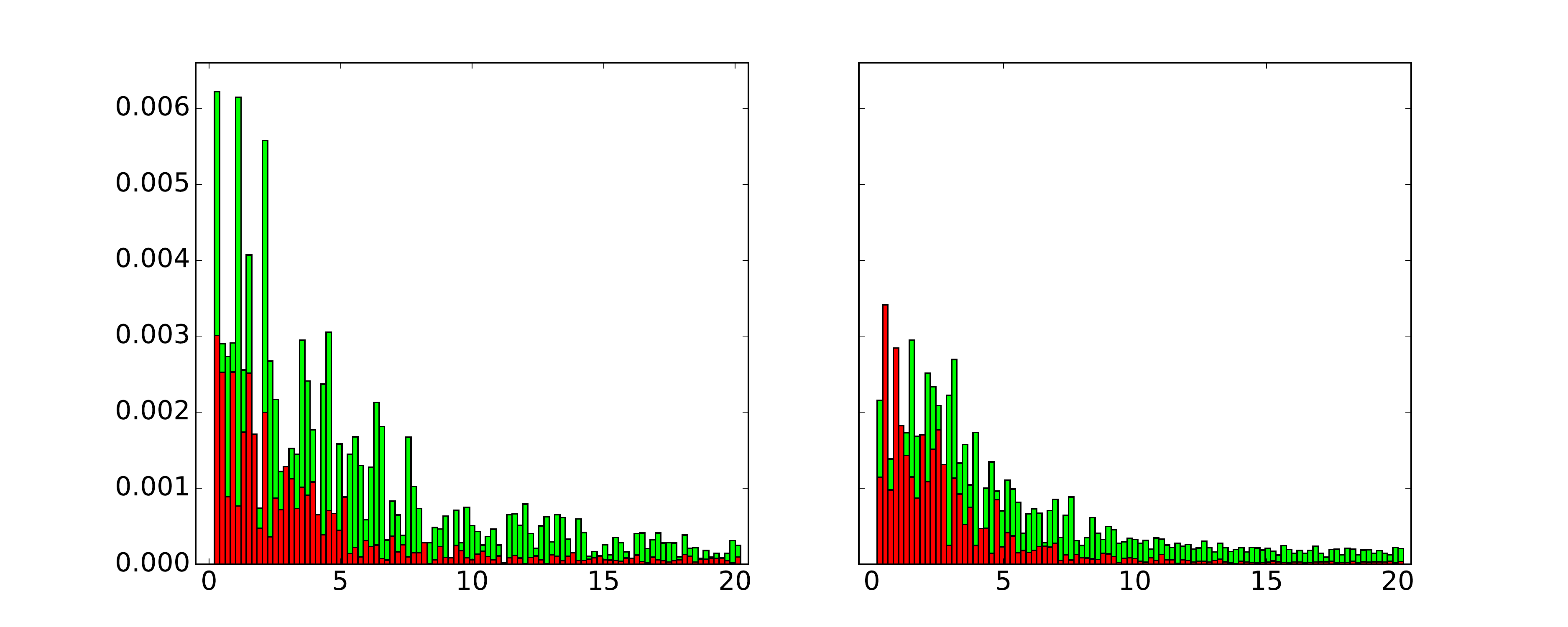}};    
\node [left of=a, xshift=-7.55cm, rotate=90] {Amplitude};
\node [below of=a, xshift=-3.8cm, yshift=-2.8cm, rotate=0] {Frequency (THz)};
\node [below of=a, xshift= 4.2cm, yshift=-2.8cm, rotate=0] {Frequency (THz)};
\end{tikzpicture}
\caption{Fourier transform of momentum-offset time series for \ce{MAPbI3} (left) and \ce{CsPbI3} (right). Fourier transforms 
are shown for valence (green) and conduction (red) band splittings. 
}
\end{figure}

\begin{figure}
\centering
\begin{tikzpicture}
\node (a) {\includegraphics[trim={0cm 0cm 0cm 0cm},clip,width=0.6\linewidth]{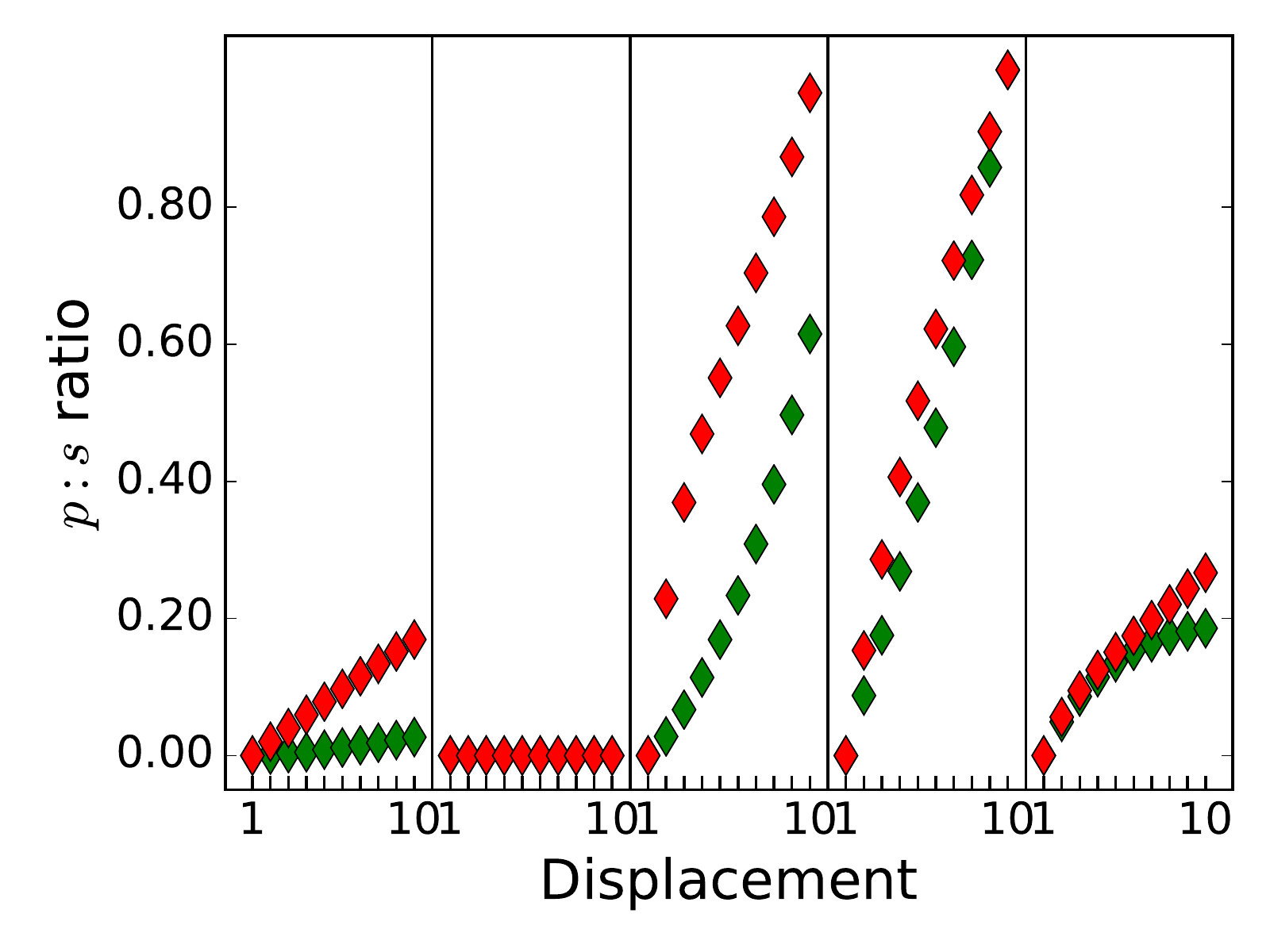}};    
\end{tikzpicture}
\caption{Ratio of $p:s$-character (green) in the valence band maximum and $p:s$-character (red) in the conduction band minimum 
versus displacement along a normal mode.
}
\end{figure}

\begin{figure}
\begin{tikzpicture}
\node (a) {\includegraphics[trim={0.0cm 0.5cm 0cm 0cm},clip,width=0.475\linewidth]{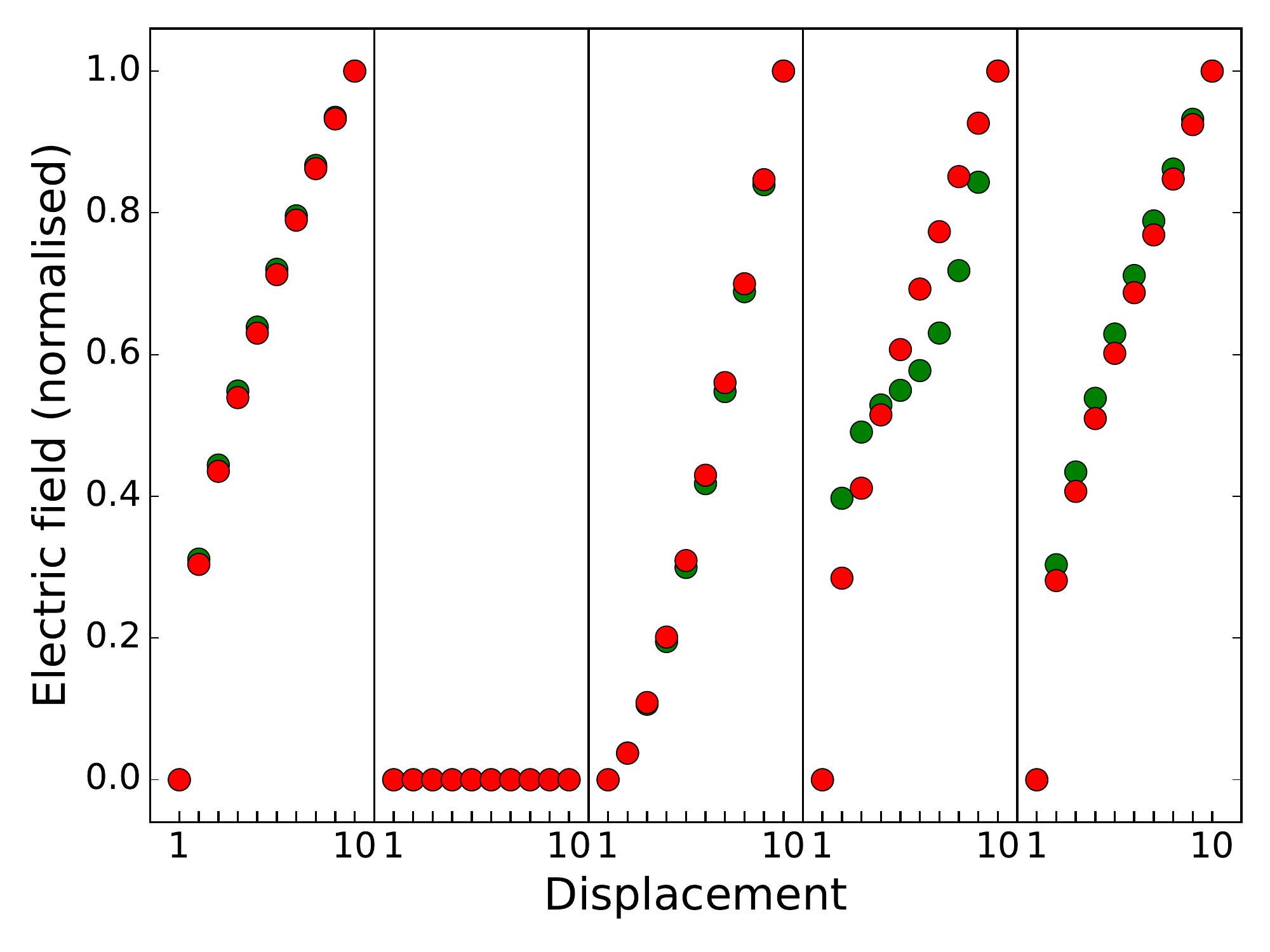}};
\node [right of =a, xshift=7.1cm] {\includegraphics[trim={0.0cm 0.5cm 0cm 0cm},clip,width=0.475\linewidth]{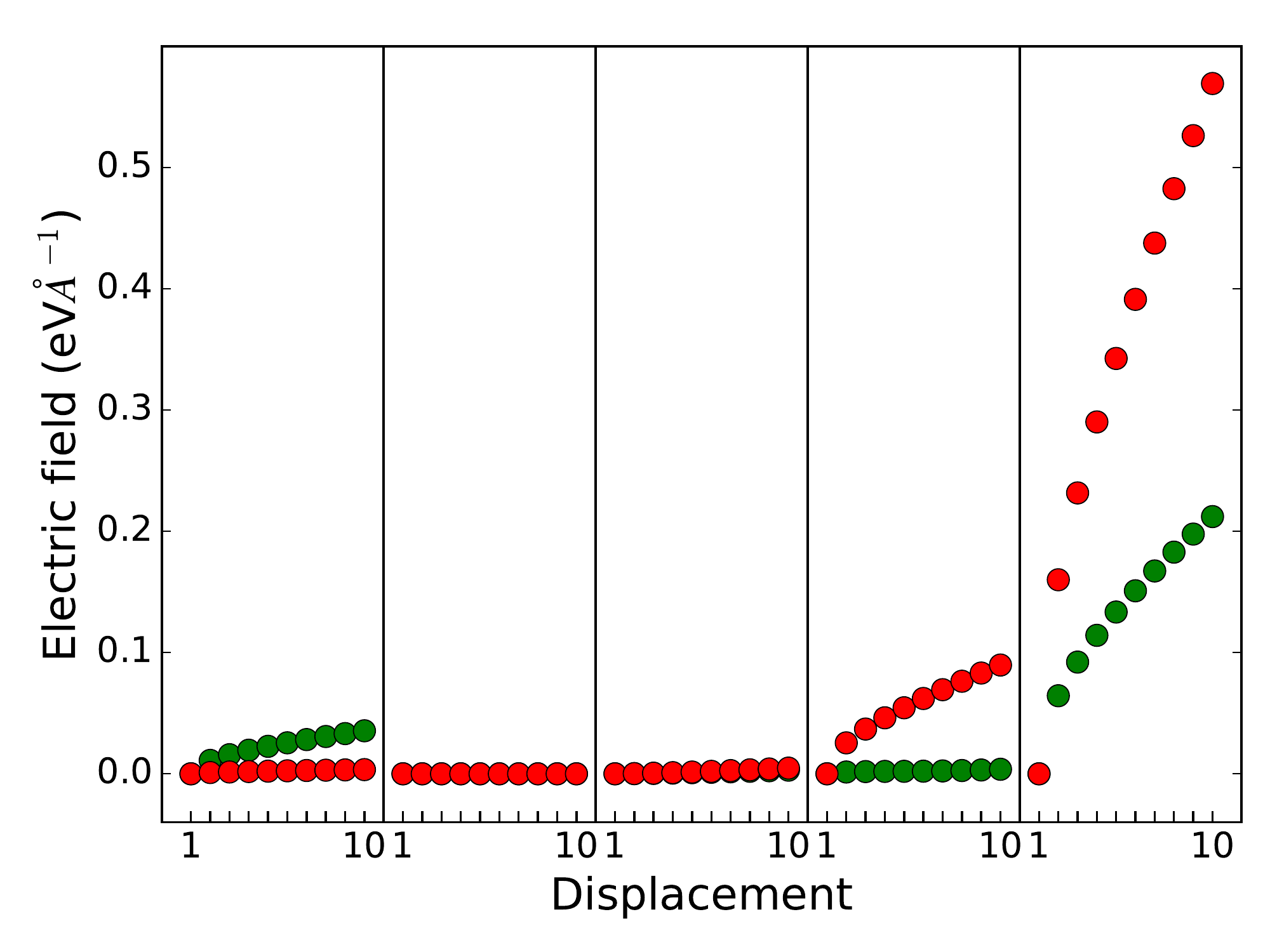}};
\end{tikzpicture}
\caption{Average electric field at the Pb site (red) and I sites (green) versus displacement along a normal mode. 
In the left handside plot, the fields have been normalised relative to the max and min in each mode. 
A vector average is taken over the three I sites. Unlike in the case of $l$ character ratios, 
the relative magnitudes of the splittings are not well described across different modes.
}
\end{figure}

\begin{figure}
\centering
\begin{tikzpicture}
\node (a) {\includegraphics[trim={3cm 4cm 2cm 5cm},clip,width=0.45\linewidth]{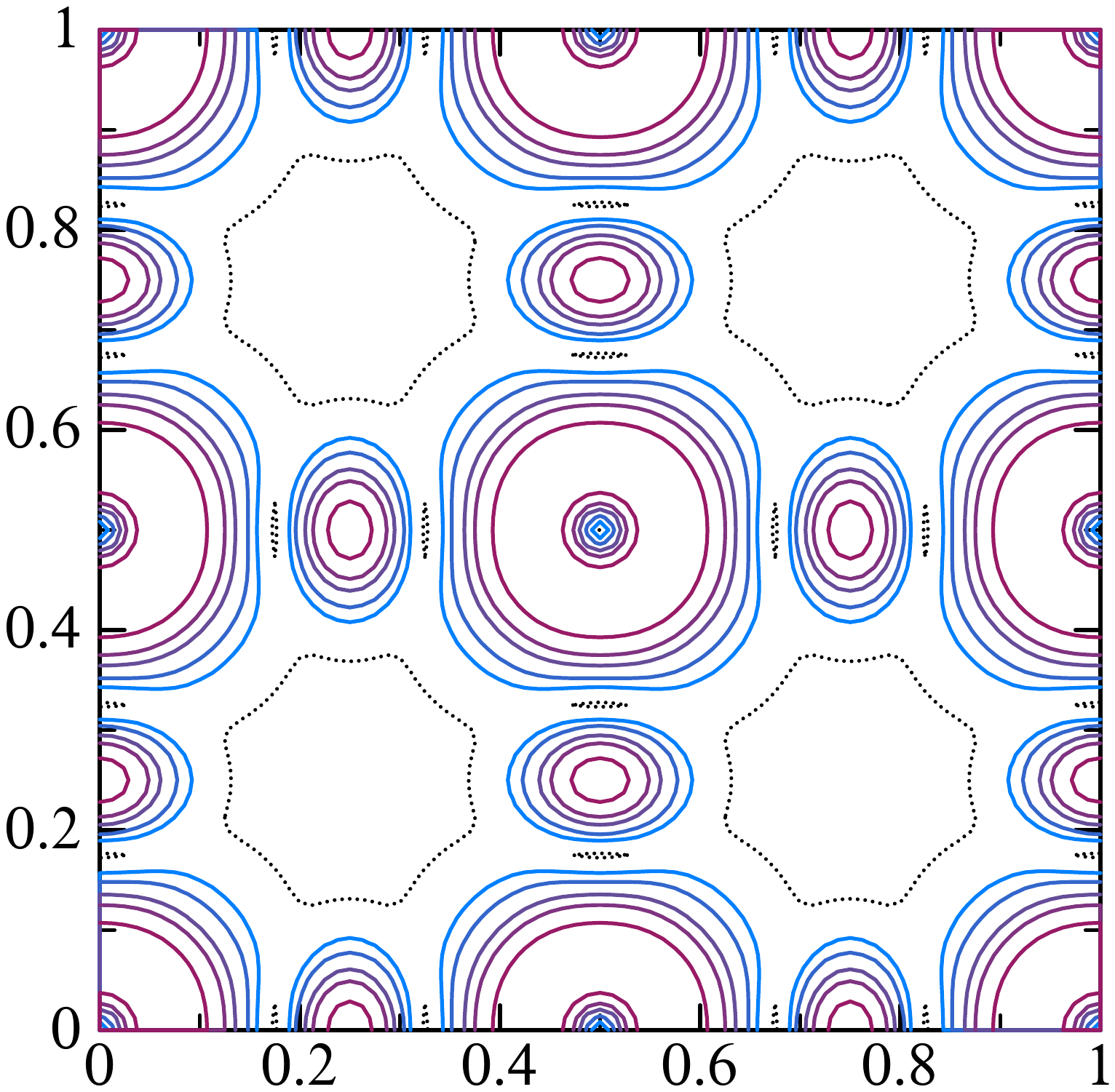}};
\node (b) [right of=a, yshift=0cm, xshift=6.5cm] {\includegraphics[trim={3cm 4cm 2cm 5cm},clip,width=0.45\linewidth]{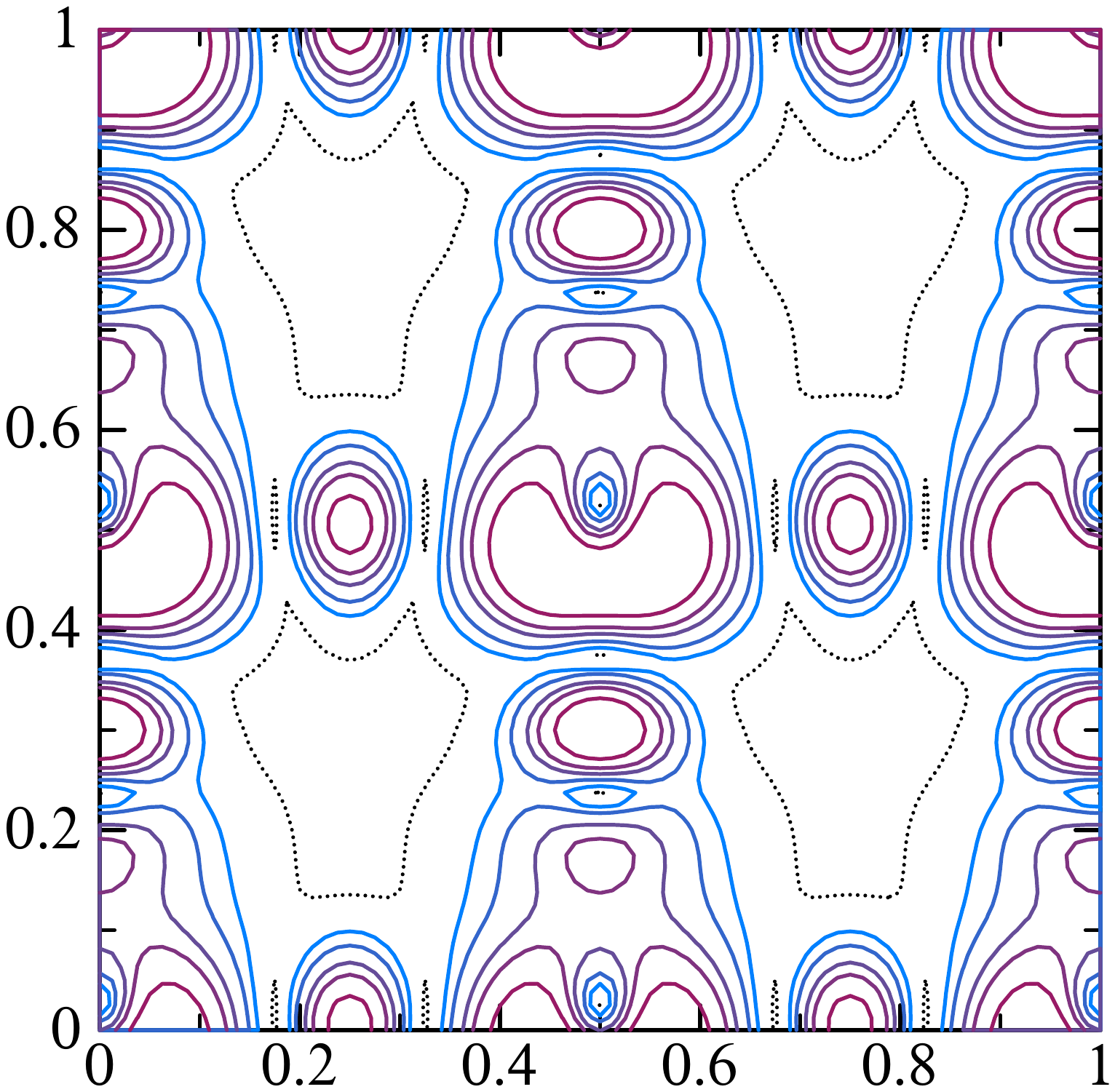}};
\node [left of=a, xshift=-2.75cm, rotate=90] {[0 0 1]};
\node [below of=a, xshift=0cm, yshift=-3.55cm, rotate=0] {[1 0 0]};
\node [below of=b, xshift=0cm, yshift=-3.55cm, rotate=0] {[1 0 0]};
\end{tikzpicture}
\caption{Example charge density plots for the undistorted (left) and max distorted (right) structures from the fifth phonon mode. 
The charge density is evaluated for the conduction band minimum eigenstates (at the $R$ point and spin-split point). A $(010)$ 
or $xz$ plane that passes through the Pb atom at the origin is shown. The cell has been doubled for better viewing, with the 
Pb atoms positioned at the corners and at the centre of the plot. 
In the distorted case, the electric field is predominantly along the $z$-axis and the momentum splitting is the the $xy$ plane.}
\end{figure}

\begin{figure}
\centering
\begin{tikzpicture}
\node (a) {\includegraphics[trim={0.0cm 0.0cm 0cm 0cm},clip,width=0.55\linewidth]{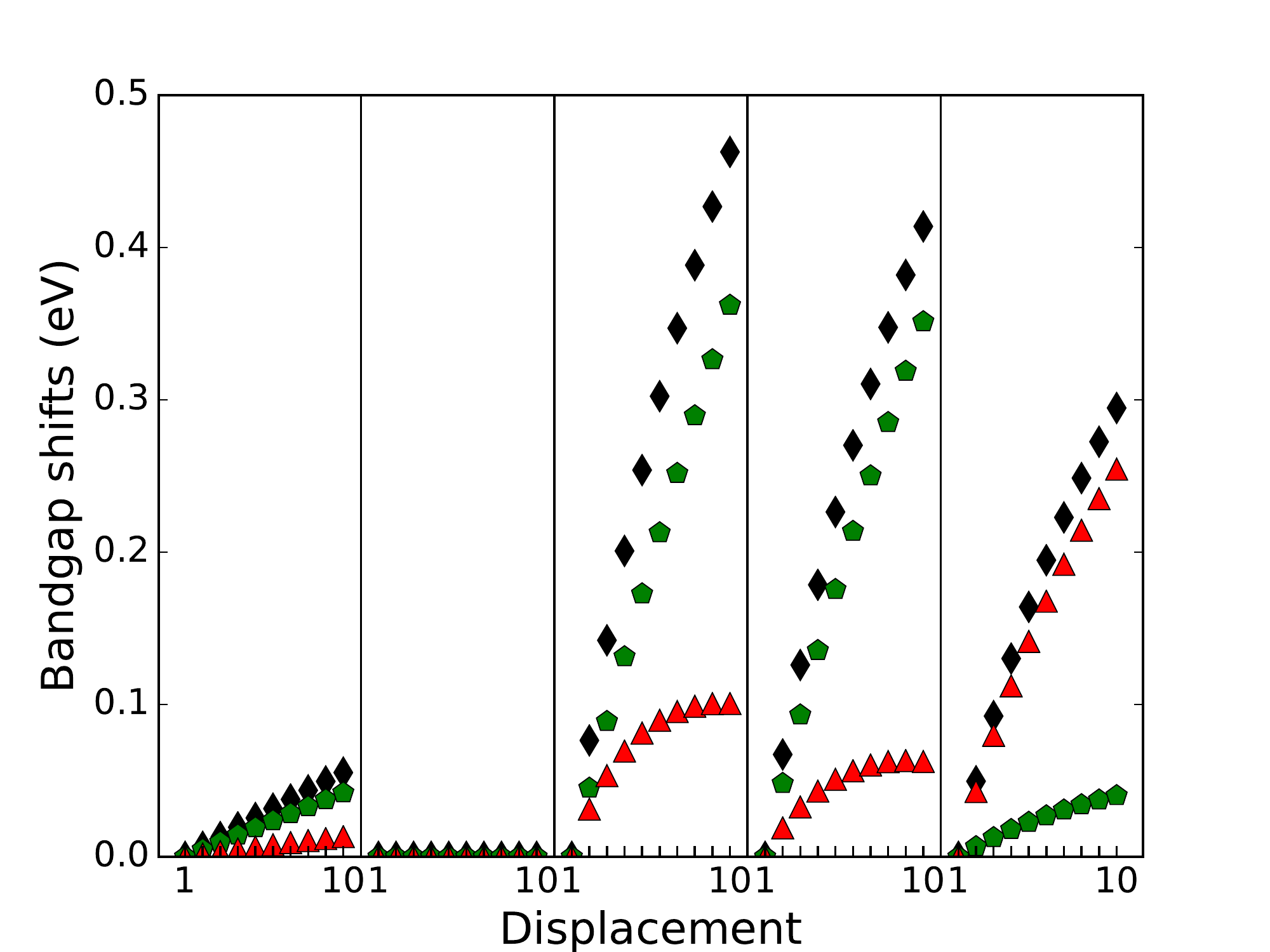}};
\node (b) [right of=a, xshift=7.95cm]{\includegraphics[trim={0.0cm 0.0cm 0cm 0cm},clip,width=0.55\linewidth]{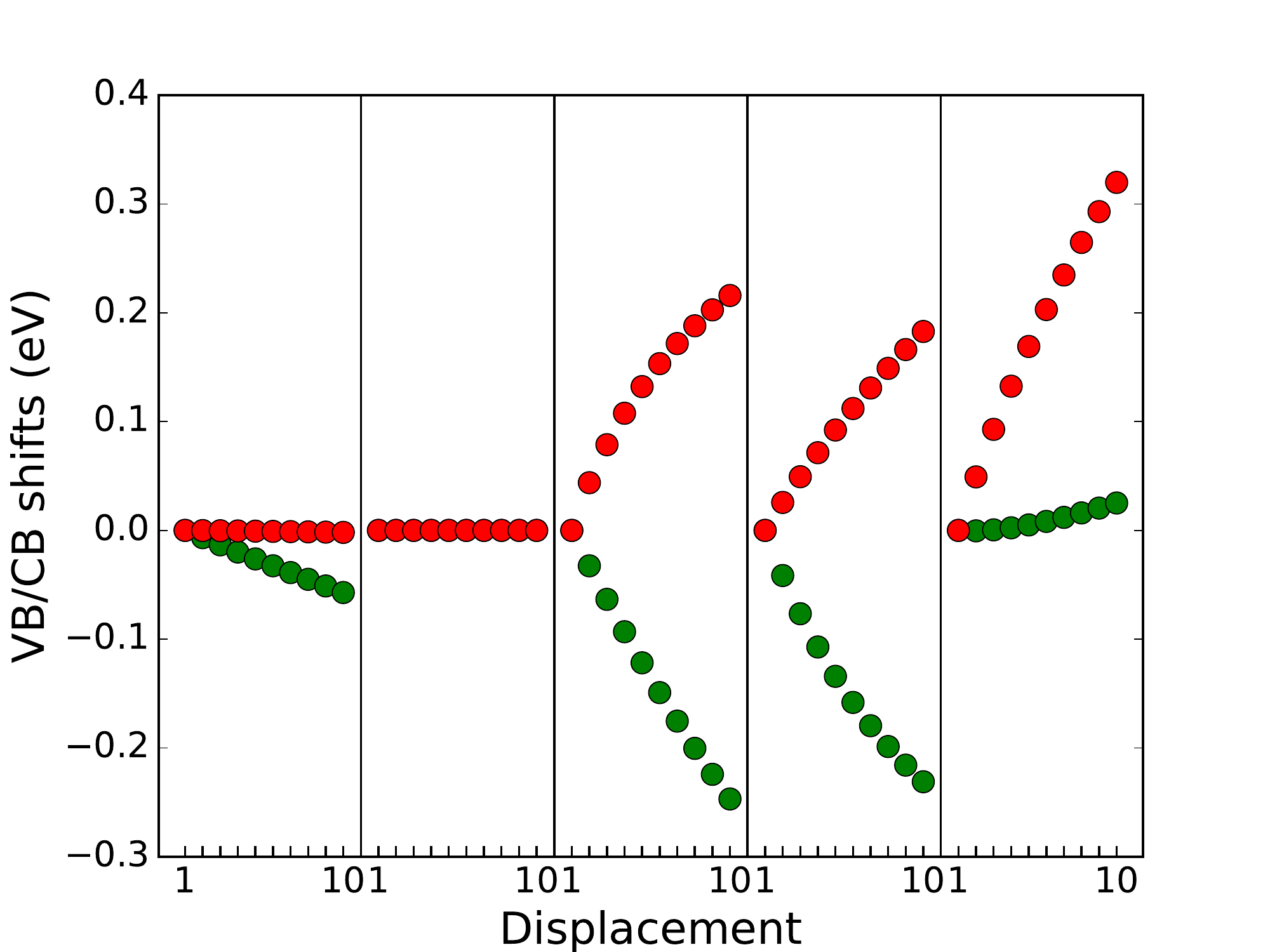}};
\end{tikzpicture}
\caption{Plot of bandgap energy changes versus displacement along a normal mode (left) and extremal point eigenvalue changes versus displacement along a normal mode (right). 
Both plots consider changes relative to the undistorted cubic structure. 
In the left handside plot, the total change with SOC is shown in black, change without SOC (scalar relativistic) is shown in 
green and change due to SOC (difference between total and scalar relativistic changes) is shown in red. In the right handside plot, 
the change in valence band maximum (VBM) is shown in green and change in conduction band minimum (CBM) is shown in red. To align the bands for different displacements, the average
potential in the unit cell is kept fixed (see Figure S10 caption). Distortion is shown to widen the gap in all cases, with a rise in 
CBM eigenvalues and tends to lower the VBM eigenvalues. Interestingly, there is a  small rise in the VBM energies for mode 5; 
possibly a result of increased antibonding for those displacements.
}
\end{figure}

\begin{figure}
\centering
\begin{tikzpicture}
\node (a) [xshift=-2cm] {\includegraphics[trim={1cm 0cm 2cm 1cm},clip,width=0.50\linewidth]{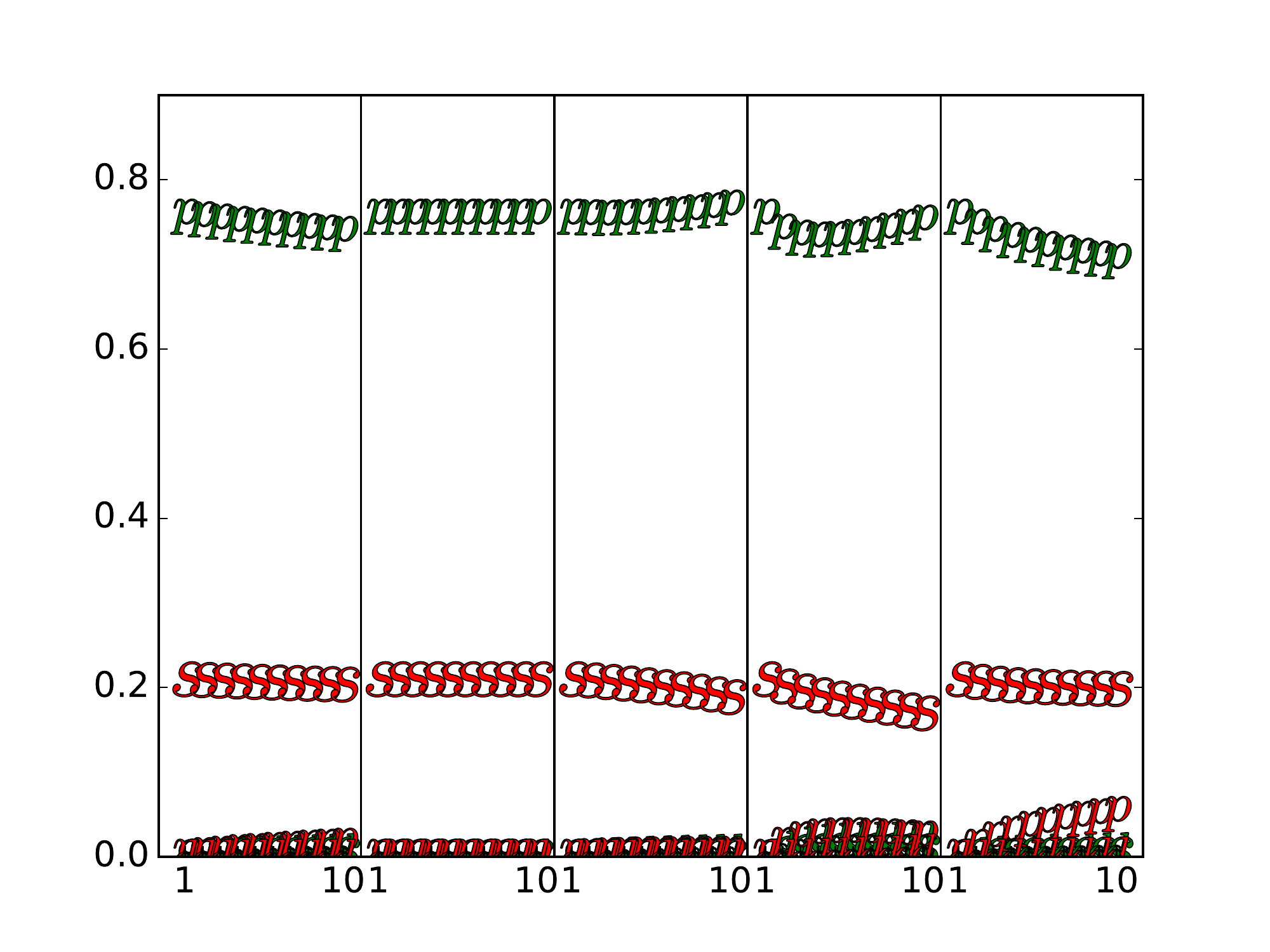}};
\node (b) [right of=a, yshift=0cm, xshift=7.5cm] {\includegraphics[trim={1cm 0cm 2cm 1cm},clip,width=0.50\linewidth]{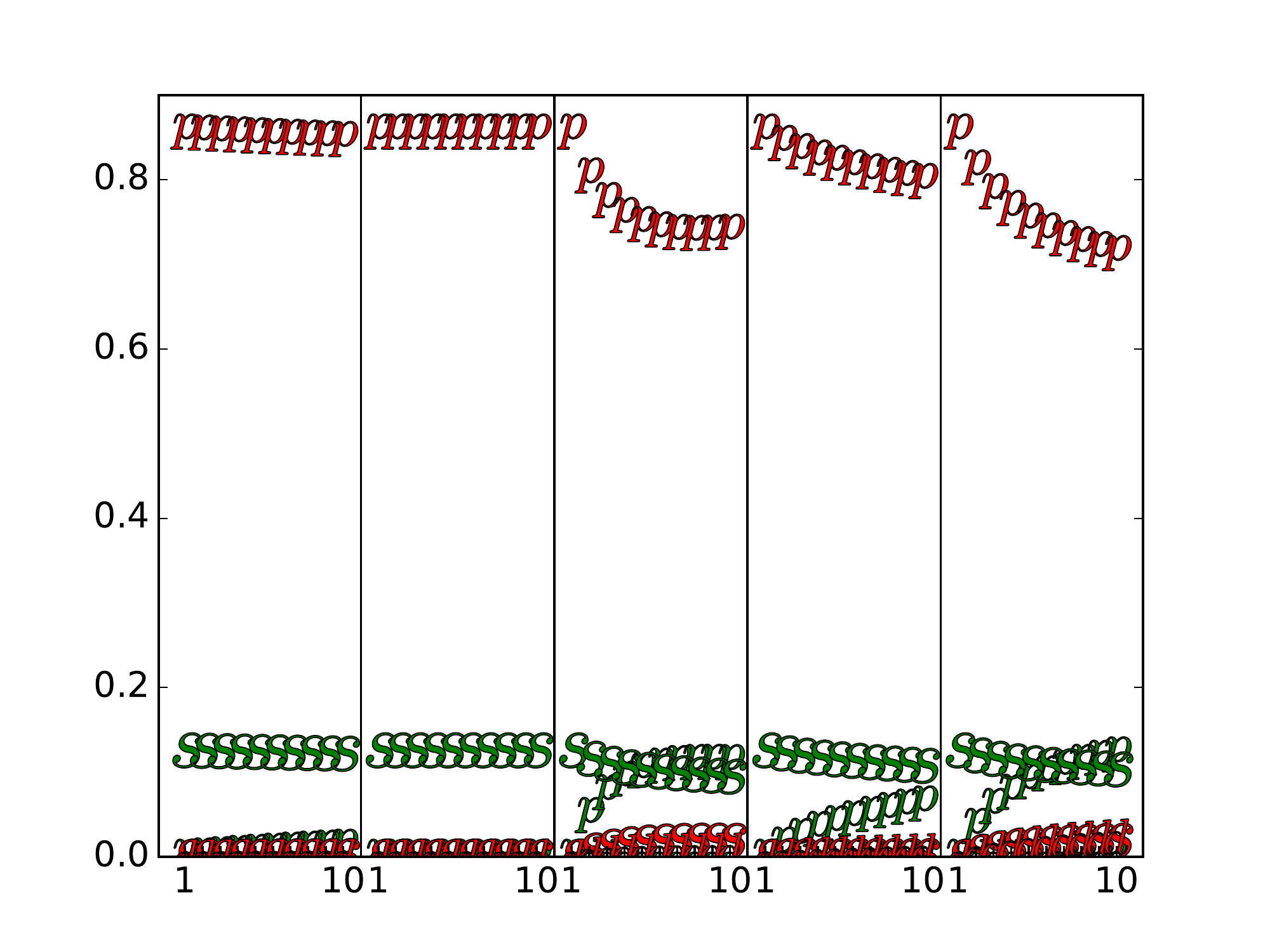}};
\node [left of=a, yshift=0cm, xshift=-3.05cm, rotate=90] {Orbital fraction};
\node [below of=a, xshift=0cm, yshift=-2.25cm, rotate=0] {Displacement};
\node [below of=b, xshift=0cm, yshift=-2.25cm, rotate=0] {Displacement};
\end{tikzpicture}
\caption{Orbital-character fraction versus displacement for valence band maximum (left) and conduction band minimum (right) eigenstates. 
The decomposition is based on a Mulliken analysis, which conveniently provides percentage orbital contributions for a given 
eigenstate. However, unlike in the partial wave analysis, the inclusion of interstitial regions means there is an ambiguity in the 
choice of basis functions in regions of overlap; there is a dependence on the basis set. The Mulliken PDOS predicts a strong I-$d$ 
character in the conduction band (right plot). However, the partial wave PDOS, which is more reliable, assigns these contributions to Pb-$p$ states. 
To account for this, the I-$d$ fraction is added to the Pb-$p$ fraction.  
Nevertheless, the basis set is constant across distorted structures and so the trends are expected to be reliable. 
The Pb and I contributions are colored red and green, while the angular momentum character $l=0$, $l=1$ and $l=2$ is represented by 
the markers $s$, $p$ and $d$. 
}
\end{figure}

\begin{figure}
\begin{tikzpicture}
\node (a) {\includegraphics[trim={1.0cm 0cm 1cm 1cm},clip,width=0.45\linewidth]{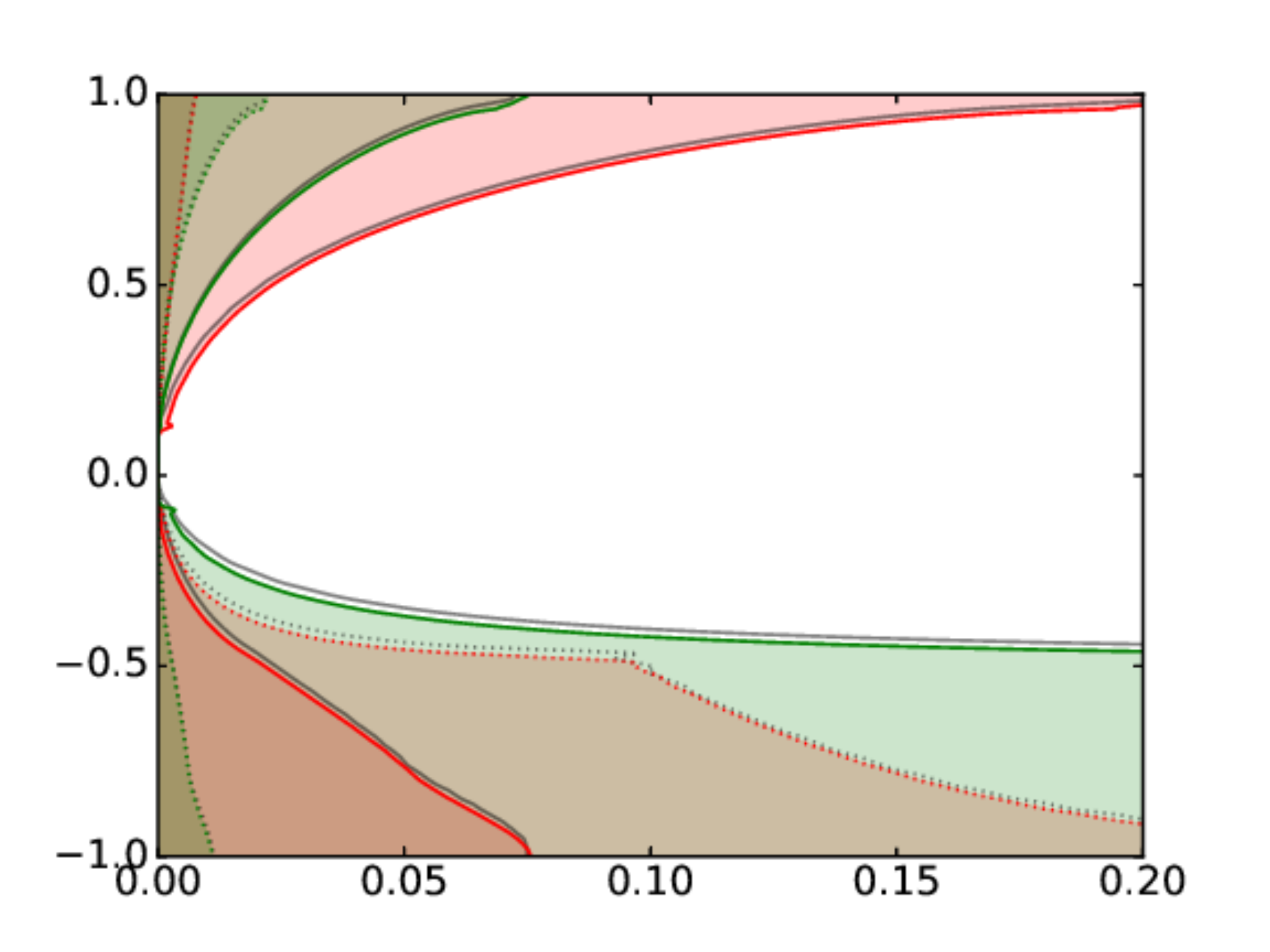}};
\node (b) [right of=a, yshift=0cm, xshift=6.5cm] {\includegraphics[trim={1cm 0cm 1cm 1cm},clip,width=0.45\linewidth]{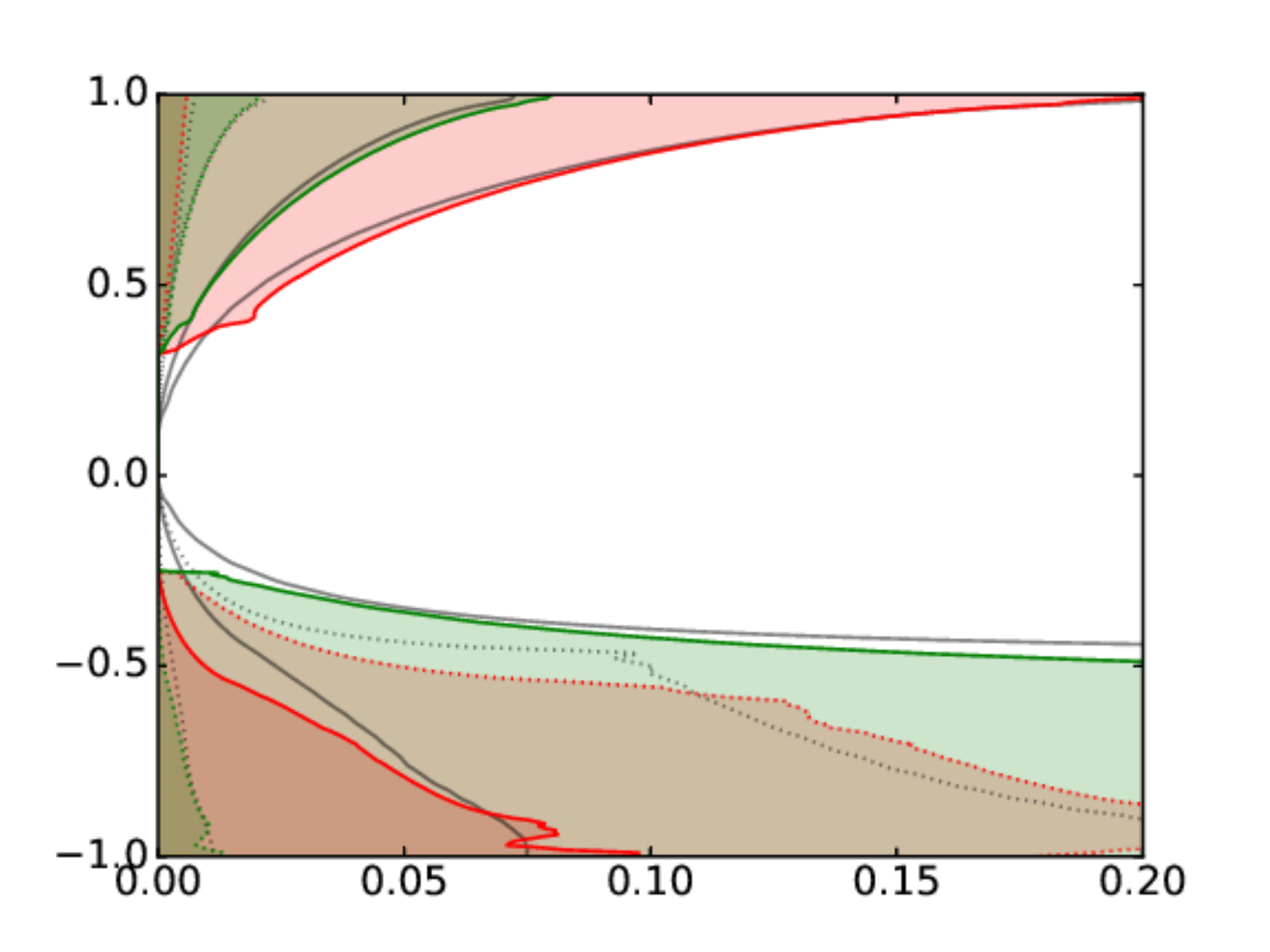}};
\node (c) [below of=a, yshift=-5cm, xshift=0cm] {\includegraphics[trim={1cm 0cm 1cm 1cm},clip,width=0.45\linewidth]{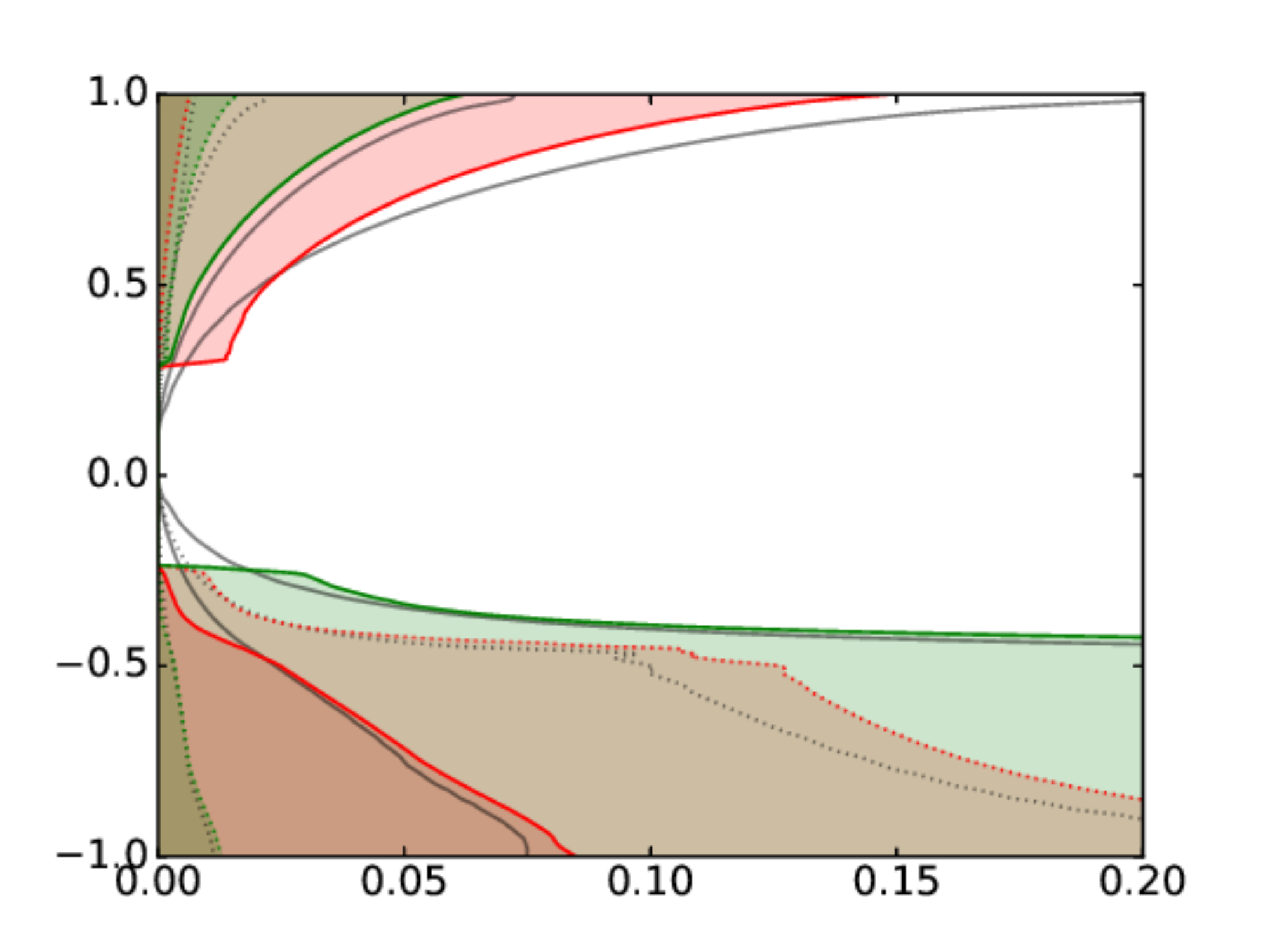}};
\node (d) [right of=c, yshift=0cm, xshift=6.5cm] {\includegraphics[trim={1cm 0cm 1cm 1cm},clip,width=0.45\linewidth]{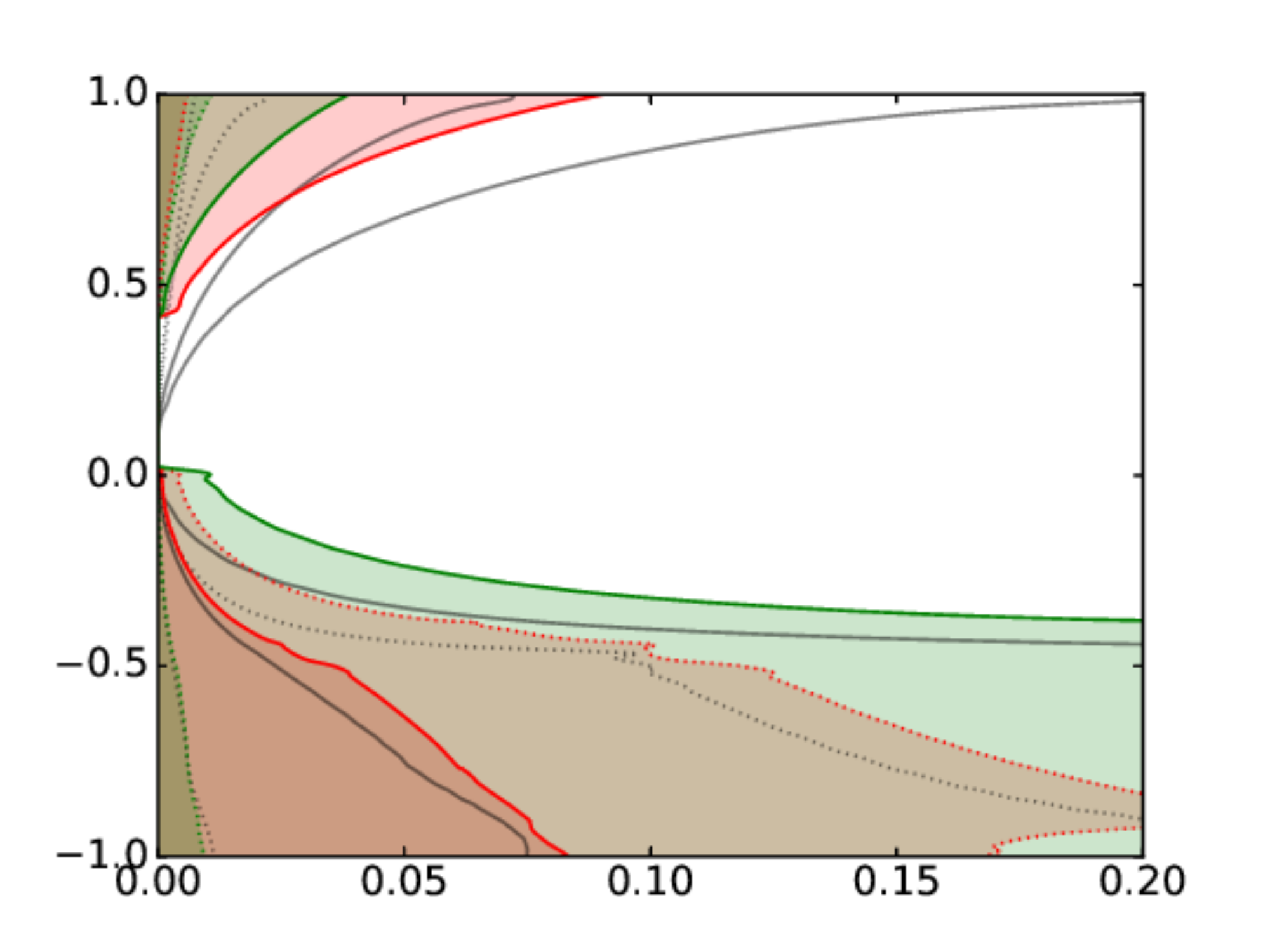}};
\node [above of=a, yshift=0.75cm, xshift=2.0cm, rotate=0] {\textbf{1}};
\node [above of=b, yshift=0.75cm, xshift=2.0cm, rotate=0] {\textbf{3}};
\node [above of=a, yshift=-5.2cm, xshift=2.0cm, rotate=0] {\textbf{4}};
\node [above of=b, yshift=-5.2cm, xshift=2.0cm, rotate=0] {\textbf{5}};
\node [left of=a, xshift=-2.95cm, rotate=90] {Energy (eV)};
\node [left of=c, xshift=-2.95cm, rotate=90] {Energy (eV)};
\node [below of=c, xshift=0cm, yshift=-1.85cm, rotate=0] {pdos};
\node [below of=d, xshift=0cm, yshift=-1.85cm, rotate=0] {pdos};
\end{tikzpicture}
\caption{LDA partial densities of states for maximum displacements along phonon modes 1, 3, 4 and 5. 
The PDOS for the reference cubic structure is shown in grey. To align the bands for different displacements, 
the average potential in the unit cell is kept fixed (this was found to be equivalent to alignment based on the 
onset of the valence I-$p$ states).
}
\end{figure}

\begin{figure}
\begin{tikzpicture}
\node (a) {\includegraphics[trim={0.0cm 0cm 0cm 0cm},clip,width=0.6\linewidth]{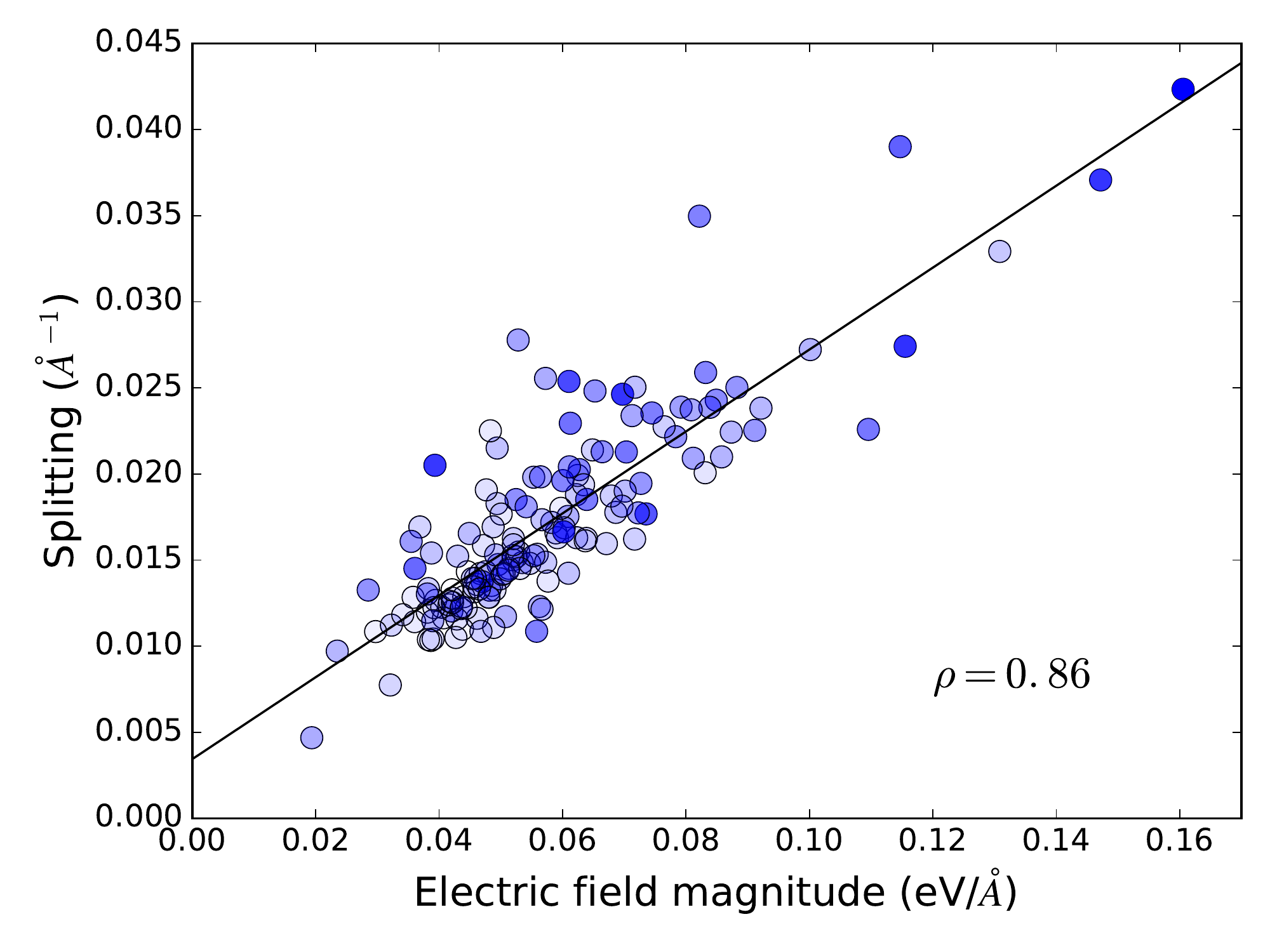}};
\end{tikzpicture}
\caption{Conduction band momentum offset plotted against the magnitude of the electric field at the Pb site for 147 stochastically sampled
MA orientations in MAPbI$_{3}$. The total energy of each structure is represented by the opacity level of the marker face colour, where higher opacity refers to higher energy. The Pearson correlation coefficient ($\rho$) between the splitting and electric field magnitudes is shown in the bottom right corner of the plot.
}
\end{figure}

\end{document}